\documentclass[]{aastex631}

\shorttitle{Preatmospheric Detection}
\shortauthors{Clark et al.}
\graphicspath{{./}{figures/}}

\newcommand{\degree}{{^{\circ}}}

\begin{document}

\title{Preatmospheric detection of a meter-sized Earth impactor}

\author[0000-0002-1203-764X]{David L. Clark}
\affiliation{Department of Earth Sciences \\
University of Western Ontario \\
London, Ontario, N6A 3K7, Canada}
\affiliation{Institute for Earth and Space Exploration \\
University of Western Ontario \\
London, Ontario, N6A 3K7, Canada}

\author[0000-0002-1914-5352]{Paul A. Wiegert}
\affiliation{Department of Physics and Astronomy \\
University of Western Ontario \\
London, Ontario, N6A 3K7, Canada}
\affiliation{Institute for Earth and Space Exploration \\
University of Western Ontario \\
London, Ontario, N6A 3K7, Canada}

\author[0000-0001-6130-7039]{Peter G. Brown}
\affiliation{Department of Physics and Astronomy \\
University of Western Ontario \\
London, Ontario, N6A 3K7, Canada}
\affiliation{Institute for Earth and Space Exploration \\
University of Western Ontario \\
London, Ontario, N6A 3K7, Canada}

\author[0000-0003-4166-8704]{Denis Vida}
\affiliation{Department of Physics and Astronomy \\
University of Western Ontario \\
London, Ontario, N6A 3K7, Canada}
\affiliation{Institute for Earth and Space Exploration \\
University of Western Ontario \\
London, Ontario, N6A 3K7, Canada}

\author[0000-0003-3313-4921]{Aren Heinze}
\affiliation{ATLAS, Institute for Astronomy \\
Honolulu, HI 96822-1839, USA}

\author[0000-0002-7034-148X]{Larry Denneau}
\affiliation{ATLAS, Institute for Astronomy \\
Honolulu, HI 96822-1839, USA}



\begin{abstract}

On 2020 September 18 US Government sensors detected a bolide with peak bolometric magnitude of -19 over the western Pacific. The impact was also detected by the Geostationary Lightning Mapper (GLM) instrument on the GOES-17 satellite and infrasound sensors in Hawaii. The USG measurements reported a steep entry angle of $67\degree$ from horizontal from a radiant $13\degree$ E of N and an impact speed of 11.7 km s$^{-1}$.  Interpretation of all energy yields produces a preferred energy estimate of 0.4 kt TNT, corresponding to a $23000$ kilogram $3$ meter diameter meteoroid. A post-impact search of telescopic images found that the ATLAS survey captured the object just 10 minutes prior to impact at an Earth-centred distance of nearly $11900$ kilometers with apparent magnitude $m\text{=}12.5$. The object appears as a $0.44\degree$ streak originating on the eastern edge of the image extending one-third of the USG state vector-based prediction of $1.26\degree$ over the 30 second exposure. The streak shows brightness variability consistent with small asteroid rotation.  The position of Earth's shadow, the object's size, and its consistency with the reported USG state vector confirm the object is likely natural. This is the eighth preatmospheric detection of a Near-Earth Asteroid (NEA) impactor and the closest initial telescopic detection prior to impact. The high altitude of peak fireball brightness suggest it was a weak object comparable in many respects with 2008 TC3 (Almahata Sitta meteorite), with absolute magnitude $H=32.5$ and likely low albedo. Therefore we suggest the NEA was a C-complex asteroid. 

\end{abstract}



\section{Introduction} \label{sec:intro}

The understanding of the Near Earth Asteroid (NEA) population brings understanding of the risks of catastrophic or locally destructive Earth-impacting events and insight into the mitigation of that risk.  Fireballs are messengers of that understanding, either acting as proxies to their larger asteroidal cousins, or as fragments of past asteroidal mixing. A rare and important opportunity arises when a meteoroid is observed in more than one of three regimes: in-space as a meteoroid, interacting with the atmosphere as a fireball, and on the ground as meteorites.   The Fireball Retrieval on Telescopic Survey Images (FROSTI) project seeks to locate serendipitous images of pre-contact meteoroids on sky survey images collected for other scientific purposes \citep{Clark2010}.  We continuously collect image borehole data from a large set of sky surveys and imaging space missions; our current catalogue comprises approximately 13.5 million images.  As newly reported and historical fireball events become available, out of atmosphere approach trajectories for fireballs are calculated and the image catalogue searched for image-trajectory intersections using a survey independent representation of the catalogue images \citep{2014PASP..126...70C}.  One such source of fireball events is the Center for Near Earth Object Studies (CNEOS) Fireball web page\footnote{\url{https://cneos.jpl.nasa.gov/fireballs/}} reporting on U.S. Government (USG) sensor data.  In parallel and referred to by CNEOS is the Bolide Detections from Geostationary Lightning Mapper (GLM) system \citep{2018M&PS...53.2445J} reporting on data collected by instruments (GLM-17 and GLM-18) on two Geostationary Operational Environment Satellite (GOES) satellites (GOES-17 and GOES-18).

\begin{figure}
\plotone{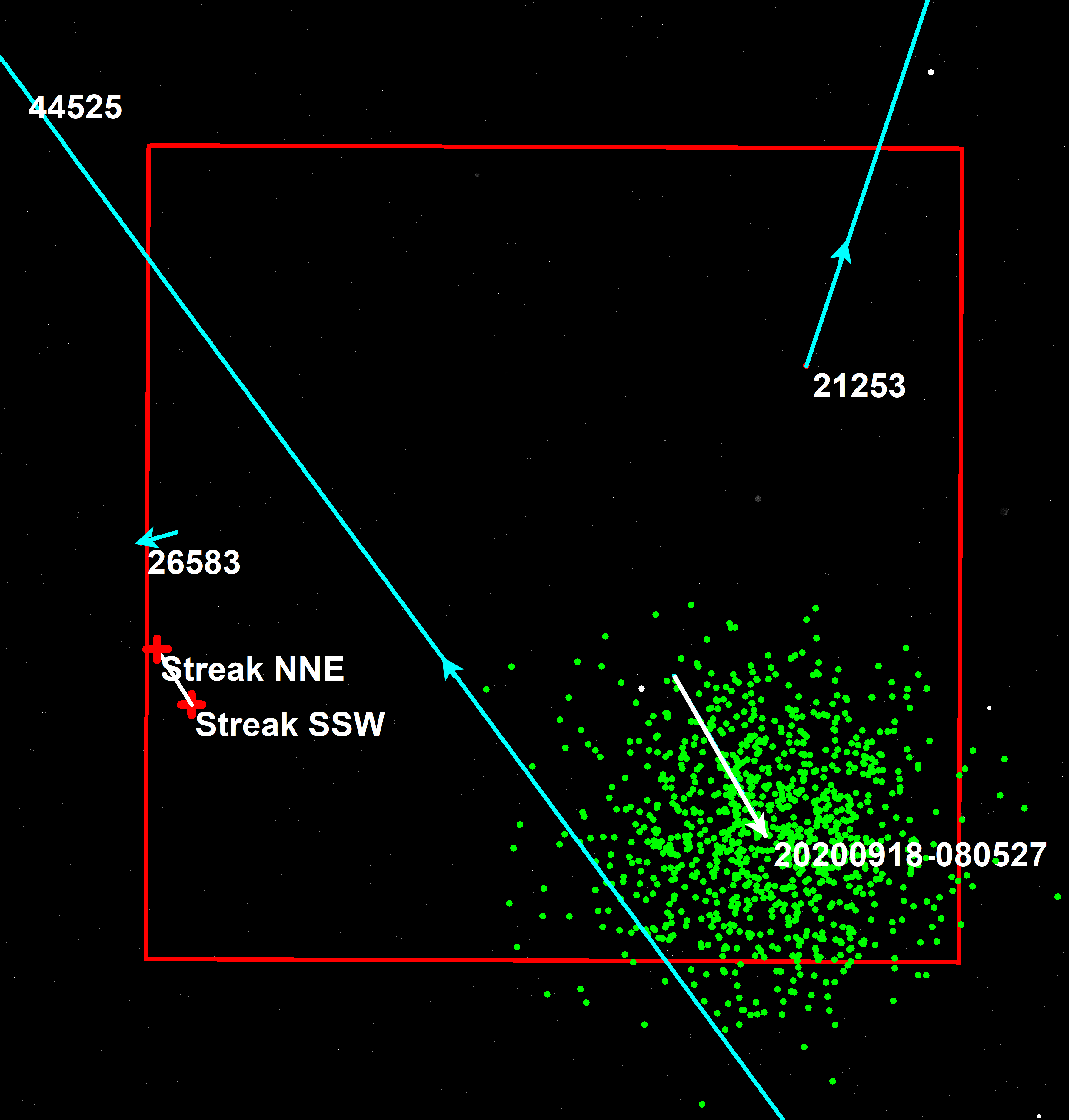}
\caption{A simulation of ATLAS Image 02a59110o0264c showing the predicted path of the USG fireball on the image based on a trajectory integration backward in time from the meteoroid contact state reported by USG.  The white arrow represents the predicted motion of the nominal object between the start and end of the exposure. Green dots represent the end of exposure positions of 1000 clones generated using a standard deviations of $0.1\degree$, $0.1$ km, and $0.1$ km~s$^{-1}$ on the USG reported values.  The position of the the actual image streak is shown in white at the left between its NNE and SSW endpoints. Shown in blue are the paths of the three catalogued satellites which crossed the field of view of the image, object 26583 appearing in the actual image in Figure \ref{fig:actualimage}, and objects 21253 and 44525 which were in the Earth's shadow and not imaged. Note the image size is 5.5$\degree$ on a side.}
\label{fig:searchimage}
\end{figure}

\begin{figure}
\plotone{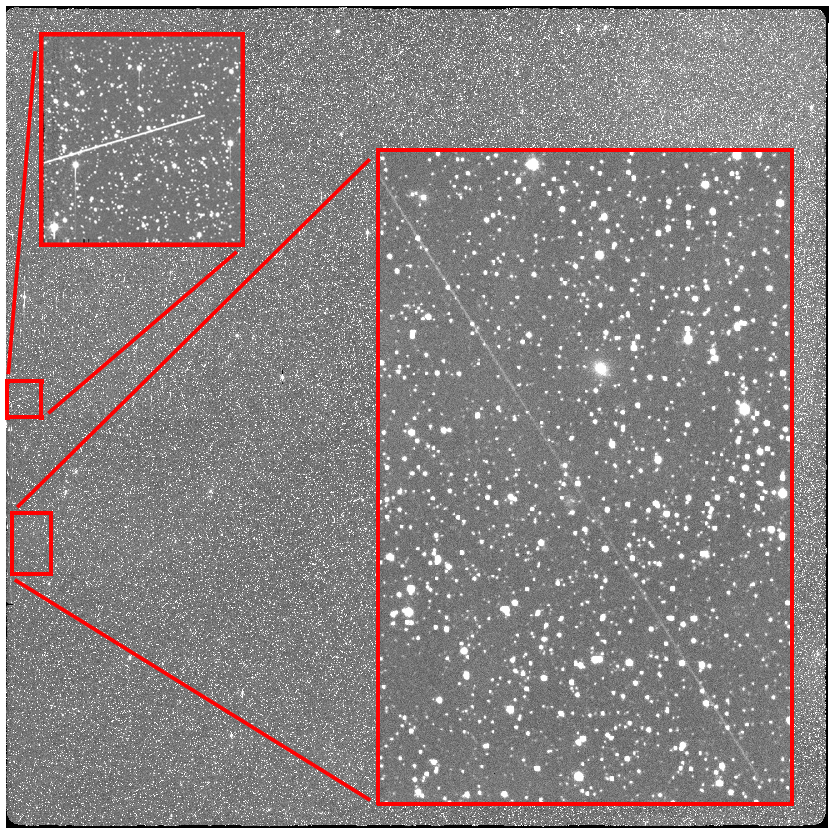}
\caption{ATLAS Image 02a59110o0264c with cut-outs showing the location of the USG 20200918 object streak (right) and the Kosmos 2319 SL-12 rocket body, both at the Eastern (left) edge of the image.} 
\label{fig:actualimage}
\end{figure}

Here we report on the first successful identification by FROSTI of a preatmospheric meteoroid.  The system identified that the meteoroid responsible for the USG 2020 September 18 fireball which occurred at 08:05:25 UT over the Pacific had potentially been imaged by the ATLAS Haleakal\a=a Telescope \citep{2018PASP..130f4505T}. The bolide was also recorded by GLM-17. A summary of the USG and GLM bolide data can be found in Tables \ref{tab:cneos} and Table \ref{tab:glm} respectively. The simulated telescopic image using the USG reported 2020 September 18 state vector to produce an estimated trailed object as created by FROSTI can be seen in Figure \ref{fig:searchimage}. The actual ATLAS image with an inset containing the trailed object linked to USG 2020 September 18 is shown in Figure \ref{fig:actualimage}. The capture was unusual in that the image was taken at 07:55:47 UT through 07:56:17 UT, only 10 minutes prior to the fireball.  This chance timing was both fortuitous and problematic. The absolute magnitude $H=32.5$ meteoroid was quite bright at apparent magnitude $m=12.5$, both due to its proximity and favourable illumination just prior to entering into Earth's shadow.  However, the object appears as an elongated streak originating off-image, the observed length representing only about one-third of the modelled motion of the object during the 30 second exposure.  Not having the complete object path on-image reduces the amount of information available for confirmation of the object's orbit, for light curve extraction to study object rotation and shape, and for predicting other sky survey images which may contain the object.   

Figures \ref{fig:searchimage} and \ref{fig:actualimage} show that the predicted location of the object based on the reported USG state vector and the observed  streak are on opposite sides of the 5.5$\degree$ image field.  This difference in position, though not unexpected for reported USG state vector data which is known to be of varying accuracy \citep{Devillepoix2019}, along with the truncated streak, requires that we justify the claim that the image is of  same object as the USG 2020 September 18 fireball.  As we will show in Section~\ref{sec:origin}, the apparent velocity of the object, its large mass (for an artificial object) together with the fact that if a preatmospheric object it is outside the Earth's shadow lead us to conclude with high confidence that the streak is the meteoroid associated with the USG fireball. 

Subsequent to the image discovery, a deeper analysis of the USG and GLM data, in particular the relatively high altitude of fragmentation, and the FROSTI-derived Apollo orbit of the object, have revealed some similarities of this object and the Almahata Sitta (2008 TC3) event.

At the time of publication there have been seven cases where Earth-impacting objects have been imaged in space (2008 TC3, 2014 AA, 2018 LA, 2019 MO, 2022 EB5, 2022 WJ1, and 2023 CX1).  Four of these events precede the USG 20200918 fireball event.  In all cases, the objects were first identified by ground-based telescopic imaging. The USG 2020 September 18 fireball event is the first case where the determination of an object's trajectory and properties are driven by the impact event, with subsequent identification of precovery images.    

\begin{deluxetable*}{lllrc}
\tablenum{1}
\tablecaption{USG Event Description\label{tab:cneos}}
\tablehead{
&&&&$\Delta$ Values for\\
&&&Best Fit&Parameter Fit \\
\multicolumn2l{Value Description} & USG Value & Value $\Delta$ & Clones
}
\startdata
Peak Brightness (UT) & Date & 2020 Sep 18 \\
& Time & 08:05:27 \\
&Peak Magnitude (Bolometric)&-19&\\
\multicolumn2l{Latitude (deg.)} & 2.4 N & -0.1509 & -0.0791$\pm$0.0774\\
\multicolumn2l{Longitude (deg.)} & 168.7 W & 0.2845 & 0.1878$\pm$0.0666\\
\multicolumn2l{Altitude (km)} & 46.0 & 0.3104 & 2.1577$\pm$7.2171\\
\multicolumn2l{Velocity (km s$^{-1}$)} & 11.7 \\
Velocity (km s$^{-1}$) & $v_x$ & 10.2 & 0.3855 & 0.2724$\pm$0.2617\\ 
& $v_y$ & 2.9 & 0.6807 & 0.6891$\pm$0.0230\\
& $v_z$ & -4.9 & -0.2369 & -0.2102$\pm$0.0362  \\
\multicolumn2l{Radiated Energy from GLM lightcurve (J)} & 5e10 \\
\multicolumn2r{Calculated Total Impact Energy (kt)} & 0.16 \\
\multicolumn2l{Radiated Energy from USG lightcurve (J)} & 5.6e10 \\
\multicolumn2r{Calculated Total Impact Energy (kt)} & 0.18 \\
\multicolumn2l{Adjusted for fireball luminous efficiency (J)} & 1.4e11 \\
\multicolumn2r{Calculated Total Impact Energy (kt)} & 0.4
\enddata
\tablecomments{First two columns: The CNEOS web page entry for the 2020 September 18 USG fireball event\footnote{\url{https://cneos.jpl.nasa.gov/fireballs/}}.  Latitude and longitude are geodetic.  Right-handed velocity components: $v_z$ is directed along the Earths rotational axis towards the north pole, $v_x$ and $v_y$ lie on the equatorial plane, with $v_x$ directed towards the prime meridian.  The kilotons of TNT Impact Energy is derived from the Total Radiated Energy by USG using an empirical expression from \citet{2002Natur.420..294B}. The altitude here corresponds to the height of peak brightness as does the velocity.  Energy values resulting from further analysis are listed for comparison (see text). The last two columns describe the change to the reported USG state vector required to best fit the candidate image streak and the USG observations. The first of these columns gives the absolute changes in values required to best fit the streak. The latter column gives the aggregate $1\sigma$ uncertainty ranges of parameter fit clones generated from the MultiNest \citep{multinest2011,pymultinest2016} parameter fit covariance matrices output in the ten runs performed,  expressed as deltas from the USG values.}
\end{deluxetable*}

\begin{deluxetable*}{lll}
\tablenum{2}
\tablecaption{GLM Bolide Event Description\label{tab:glm}}
\tablehead{\multicolumn2l{Value Description} & Value}
\startdata
Bolide & Date/Time (UT) & 2020 Sep 18 08:05:25 \\
& Latitude (deg.) & 2.3 \\
& Longitude (deg.) & -169.9 \\
& Detected by & GLM-17 \\
& How Found? & algorithm \\
& Other Detecting Sources & USG \\
& Confidence Rating & high \\
Signal & Start Time (UT) & 2020 Sep 18 08:05:25.675 \\
& End Time (UT) & 2020 Sep 18 08:05:27.019 \\
& Duration (seconds) & 1.344 \\
& Latitude (deg) & 2.5 \\
& Longitude (deg) & -169.9 \\
& Total radiated energy (J) & 5e10 
\enddata
\tablecomments{The GLM web page entry for the 2020 Sep 18 event\footnote{\url{https://neo-bolide.ndc.nasa.gov/}}}.
\end{deluxetable*}

\section{Object Detection and Image Search} \label{sec:detection}
The FROSTI project \citep{Clark2010} searches for preatmospheric observations of meteoroids through the backwards numerical integration of measured state vectors  of fireball events and compares look angles to  an ever-growing collection of sky survey images.  FROSTI's current database contains over 13.5 million image boreholes from both ground-based telescopes and spacecraft.  Each fireball state vector is integrated back in time using the RADAU integrator \citep{1985ASSL..115..185E}. A 1000-clone uncertainty cloud is generated for each event based on state vector uncertainties.  In cases where multiple trajectory or physical models lead to differing state vectors,  each model is treated as a separate event. Results from all such fireball scenarios and images yield a list of potential precovery images that are then prioritized for manual visual inspection based on a predicted object's apparent magnitude, the portion of the uncertainty cloud on a given image, and the limiting magnitude of the image in question.  From over a decade of FROSTI monitoring the 94\% probability of an 11.5 magnitude USG fireball of 2020 September 18 appearing on ATLAS 1 image 02a59110o0264c taken 10 minutes prior to the fireball event was exceptional (see Figure \ref{fig:searchimage} for the simulated image capture, regenerated after the initial find to include a 10000 clone uncertainty cloud).  Based on the nominal USG contact state vector the pre-impact meteoroid should have been seen as a 4519 arcsec long streak on the East (right) side of the image.  Later measurement of the actual streak resulted in a visual magnitude of 12.5, very similar to FROSTI's prediction.

Table \ref{tab:cneos} and Table \ref{tab:glm} summarize the bolide as reported by USG and GLM. Slight differences in the impact locations between USG and GLM are due to GLM's assumed (and fixed) height of 16 km near the equator \citep{Jenniskens2018}. The GLM lightcurve produced excellent agreement in total radiated energy with the USG value, being only  10\% less than that recorded by USG.        

In general, FROSTI determines object visibility on an image from a calculated object size expressed by diameter or absolute magnitude.  The estimated visibility of a USG reported object is arrived at by calculating kinetic energy from impact energy using the formula documented on the CNEOS Fireballs Introduction web page\footnote{\url{https://cneos.jpl.nasa.gov/fireballs/intro.html}}, converting to a mass and diameter assuming a bulk density of $3000$~kg~m$^{-3}$, and using the standard absolute magnitude methodology of \citet{1989aste.conf..524B}.  For the initial search, this produced an initial estimate of mass $M$ = $8600$ kg, diameter $d$ = $1.8$~m, and absolute magnitude $H$ = $31.5$.   Measurement of the image streak and interpretation of the GLM and USG light curves applying more physical luminous efficiency values at low entry speeds (described in Section~\ref{sec:physical}) produced a final best estimate of the object as having $H$ = $32.5$. This refined and fainter absolute magnitude actually corresponds to higher diameter and mass based on an assumed lower bulk density (similar to 2008~TC3) resulting in a correspondingly lower albedo than usually assumed by FROSTI.

\begin{figure}
\plotone{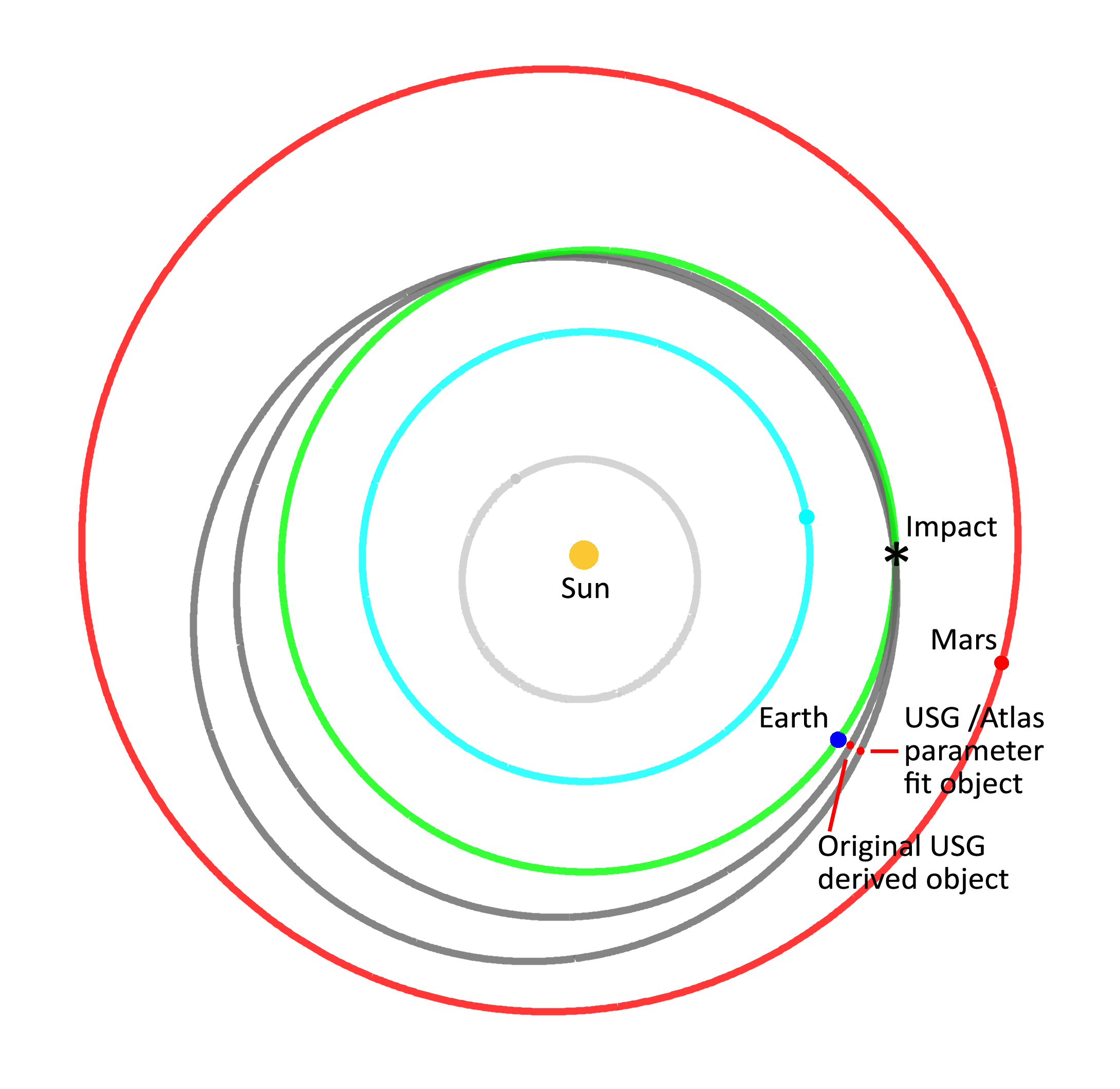}
\caption{Illustration of the meteoroid's Apollo Earth-crossing orbit. The inner meteoroid orbit is that derived directly by FROSTI from the USG event for use in the search for precovery images.  The outer orbit is the refined orbit computed from the best fit trajectory using both the USG observation and the object streak on the ATLAS image. The planets and meteoroids are positioned one month prior to the fireball event.  The impact point of Earth and the meteoroid is marked with the asterisk.}
\label{fig:orbitdiagram}
\end{figure}

Back-integration of the contact state vector and uncertainties for a period of two months yields the pre-Earth encounter orbit given in Table \ref{tab:elements} and shown in Figure \ref{fig:orbitdiagram}.  The meteoroid was in an Apollo orbit with Earth impact occurring on the inward leg of the object's orbit, making its approach a nighttime one favourable for telescopic observation. The near Earth-like orbit resulted in a low Earth-relative velocity at encounter.        

The FROSTI image search system generates simulations of object-image intersections, an example of which can be seen in Figure \ref{fig:searchimage}, and a large amount of nominal object and uncertainty clone motion information, highlights of which can be found in Table \ref{tab:image}.  The expected scenario in developing the search system was that that there would be a set of similarly pointed images all or most of which would contain an object at varying positions.  The nominal object and uncertainty clone motion analysis would be performed across the set of images, with manual image blinking or automated object searching then being performed, much as typical near-Earth object detection proceeds.  However, in the case of this object where the epoch of the capturing image is just 10 minutes prior to object impact, the on-image motion of the object is immense (4519 arcsec or $1.25\degree$ in 30 s). The image was part of a quartet, but with such extreme motion the meteoroid was outside the fields of the other quartet images.   The initial image search was performed with the standard 1000-clone uncertainty cloud, with the search result indicating that the object had a 99.9\% probability of being on the image at the beginning of the exposure and a 93\% probability at the end.  The object was predicted to appear as a streak on the East (right) side of the image extending in a SSW direction.  

The \textbf{USG reports do not include} uncertainties in position and velocity, so as a default we use standard deviations of one reported decimal place on each state element ($0.1\degree$, $0.1$ km, and $0.1$ km~s$^{-1}$) and a standard deviation of 1.0 s on the reported contact time to determine on-image probability.  Using the standard calculation for apparent magnitude $m$ as documented in \citep{1989aste.conf..524B} and an assumed phase angle slope parameter ($G$) of 0.15, we calculate $m=12.4$ representing an unusually bright candidate for detection.  However, the extreme on-image motion was expected to introduce substantial trailing loss, estimated to be on the order of nine magnitudes.  

USG fireball reports have been found to have large uncertainties in velocity \citep{Brown2015,Devillepoix2019}. To address the possibility that the chosen uncertainties of the USG reported values may be too small and may have disqualified other images potentially containing the object, we performed an additional image search with the standard deviation expanded three-fold to identify any images competitive with the proposed capture image.  The proposed capture image object intersection probability was reduced to 78$\%$ not unexpectedly, but the new uncertainty cloud did now incorporate the candidate streak.  Only one additional image was found with an object intersection probability greater than 5$\%$ at a time when the object would have been brighter than an image’s limiting magnitude and no streak was visible on that image.

Figure \ref{fig:actualimage} is the actual image ATLAS 1 image 02a59110o0264c. The image is centred on right ascension 20$^h$6$^m$30$^s$ and declination 8$\degree$10'41".  The image field is approximately $5.5\degree$ per side with celestial North up and celestial East to the right. Visual inspection of the image revealed an illuminated 1590 arcsec (0.44$\degree$) streak, one end located at the West (left) edge of the image, the other end on image (see Figure \ref{fig:actualimage} inset).  It appears almost exactly 4$\degree$ east of the predicted position and 0.9$\degree$ to the south.  The streak's slope is only 2.1$\degree$ steeper than predicted. The overall length of the streak and its direction of travel (SSW or NNE) cannot be determined from the image alone. A comparison of the predicted position of the streak from FROSTI as compared to the observed is provided in Table \ref{tab:streak}.    

The 4$\degree$ offset between the predicted object position and the imaged position highlights that it was fortuitous the manual image identification was made.  A small offset in USG contact state or in the actual orbit of the object could have resulted in the position difference being more than one field of view, substantially lessening the chance the image would have been identified and the event investigated.  The position difference also highlighted that the physical object could possibly appear on other as yet unidentified images which did not contain the FROSTI-predicted object.

\begin{deluxetable*}{lllll}
\tablenum{3}
\tablecaption{Assumed and Derived Meteoroid Properties\label{tab:properties}}
\tablehead{Property Description & & Original & & Refined}
\startdata
Total Radiated Energy (J) & Observed & 4.1E10 & Observed  & 1.4E11 \\
Impact Energy (kt) & Derived & 0.14 & Derived &  0.4 \\
Velocity (km s$^{-1}$) & Observed & 11.7 & Observed  & Same \\
Mass (kg) & Derived & 8600 & Derived & 23000 \\
Absolute Magnitude (H) & Derived & 31.5 & Derived & 32.5\\
Density (kg m$^{-3}$) & Assumed & 3000 & Derived & 1660\\
Diameter (m) & Derived & 1.8 & Derived & 3\\ 
Albedo & Assumed & 0.15 & Derived  & 0.03\\
Slope Parameter & Assumed & 0.15 & Assumed & Same
\enddata
\tablecomments{Summary of the derived vs. directly observed object attributes. These were used both for calculating on-image object appearance and in determining if the object was natural or man-made. The original USG total radiated energy was used for estimating the object mass and therefore diameter in the original FROSTI image search.  The standard FROSI bulk density of 3000 kg~m$^{-3}$ and albedo of 0.15 were used.  Post analysis including the GLM and infrasound observations together with revised luminous efficiency and similarity to 2008~TC3 yields a best estimate 3-fold greater mass.  Analysis of the presumed object streak light curve revealed that initial FROSTI magnitude estimates were one magnitude too bright.  High altitude fragmentation is consistent with a much lower bulk density. A very low albedo is required in order to enforce consistency of the derived diameter and the apparent magnitude.}
\end{deluxetable*}

\begin{deluxetable*}{ll}
\tablenum{4}
\tablecaption{Image and Predicted Object Appearance\label{tab:image}}
\tablehead{Description & Value}
\startdata
Sky Survey & ATLAS 1 \\ 
Image ID & 02a59110o0264c \\
Start Time (UT) & 2020 Sep 18 07:55:47 \\ 
End Time (UT) & 2020 Sep 18 07:56:17 \\
Right Ascension & 20$^h$6$^m$30$^s$.730 \\
Declination & 8$\degree$10'41.520"  \\
Exposure (sec) & 30 \\
On Image Probability (Start) & 99.9\% \\
On Image Probability (End) & 93.1\% \\
On image Motion (arcsec) & 4519.2 \\
Phase Angle (deg) & 55.8 \\
Apparent Magnitude (m) & 12.4 \\
Trailing Loss (Magnitudes) & 9.1
\enddata
\tablecomments{The on-image probabilities are based on a 1000 clone uncertainty cloud gravitationally integrated from a set of contact states, being the mean USG reported state with standard deviations of one reported decimal place on each state element ($0.1\degree$, $0.1$ km, and $0.1$ km~s$^{-1}$).  A standard deviation of 1.0 s is used over the USG reported impact time.}
\end{deluxetable*}

\begin{deluxetable*}{llll}
\tablenum{5}
\tablecaption{Orbital Elements\label{tab:elements}}
\tablehead{Element & Element Description & USG & Best Fit}
\startdata
 a & Semi-major axis (au) & $1.077\pm0.013$ & $1.117\pm0.034$ \\
 1/a & Semi-major axis reciprocal (au $^{-1}$) & $0.930\pm0.011$ & $0.896\pm0.027$ \\
 q & Perihelion (au) & $0.965\pm0.002$ & $0.948\pm0.003$ \\
 Q & Aphelion (au) & $1.189\pm0.028$ & $1.285\pm0.071$ \\
 e & Eccentricity & $0.103\pm0.012$ & $0.150\pm0.028$ \\
 i & Inclination (deg) & $4.968\pm0.463$ & $6.530\pm0.787$ \\
 $\Omega$ & Longitude of Ascending Node (deg) & $176.15\pm0.16$ & $175.843\pm0.197$  \\
 $\omega$ & Argument of Perihelion (deg) & $234.52\pm2.32$ & $236.145\pm4.026$ \\
 f & True anomaly (deg) & $-110.18\pm2.62$ & $-110.108\pm4.618$\\
 M & Mean anomaly (deg) & $261.23 \pm4.05$ & $266.663\pm8.073$ \\
 Tp & Time at Perihelion (TD) & 2020 Nov 9 06:39:16 & 2020 Nov 8 16:46:49 \\
 & ($\pm$ in days) & $\pm2.612$ & $\pm4.586$\\
 Tj & Tisserand's Parameter & $5.7\pm0.05$ & $5.574\pm0.131$ \\
 & Epoch (TD) & 2020 Jul 20 08:06:36 & 2020 Jul 20 08:06:36
\enddata
\tablecomments{USG: Orbital elements of the object determined by back-integrating the reported USG contact state in Table \ref{tab:cneos} for two months, using an uncertainty cloud of one significant digit in the state vector components as standard deviation ($0.1\degree$, $0.1$ km, and $0.1$ km s$^{-1}$).  Best Fit: Orbital elements determined by back-integrating the 20000 clones produced by ten parameter fitting runs finding the best self-consistent match  of the on-image streak and USG observations as described in Section \ref{sec:combining}.} 
\end{deluxetable*}

\begin{deluxetable*}{lrrlrl}
\tablenum{6}
\tablecaption{Object Streak\label{tab:streak}}
\tablehead{&Predicted & On-image  & & Difference&}
\startdata
NNE end-point RA (deg)  & 300.807 & 304.319 & \kern-1em*\\
NNE end-point Dec (deg) & 7.359 & 7.516 &  \kern-1em*\\
SSW end-point RA (deg) & 300.168 & 304.080 & & 3.912 & \\
SSW end-point Dec (deg) & 6.276 & 7.142 & & 0.866 & \\
Angular difference at SSW point (deg) & & & & 3.981 & \\
Streak $\Delta$ RA (deg) & -0.369 & -0.239 &  \kern-1em* & 0.130 &  \kern-1em* \\
Streak $\Delta$ Dec (deg) & -1.083 & -0.373 &  \kern-1em* & 0.710 &  \kern-1em* \\
Streak length (deg) & 1.255 & 0.442 &  \kern-1em* & -0.813 &  \kern-1em*\\
Slope counter clockwise from W (deg) & 120.546 & 122.647 & & 2.101 & \\
\enddata
\tablecomments{Comparison of predicted nominal object streak based on USG observation and observed NNE-truncated streak on the actual image. *Value impacted by edge truncation.} 
\end{deluxetable*}

\section{Fireball Properties} \label{sec:fireball}

As the impact occurred over open ocean, all data available for the associated fireball are from three sources: US Government sensors\footnote{\url{https://cneos.jpl.nasa.gov/fireballs/}}, the Geostationary Lightning Mapper on-board the Geostationary Operational Environment Satellite (GOES) 17 stationed over the Pacific, and an infrasound array located in Hawaii.

\subsection{Energy}
\begin{figure}
\plotone{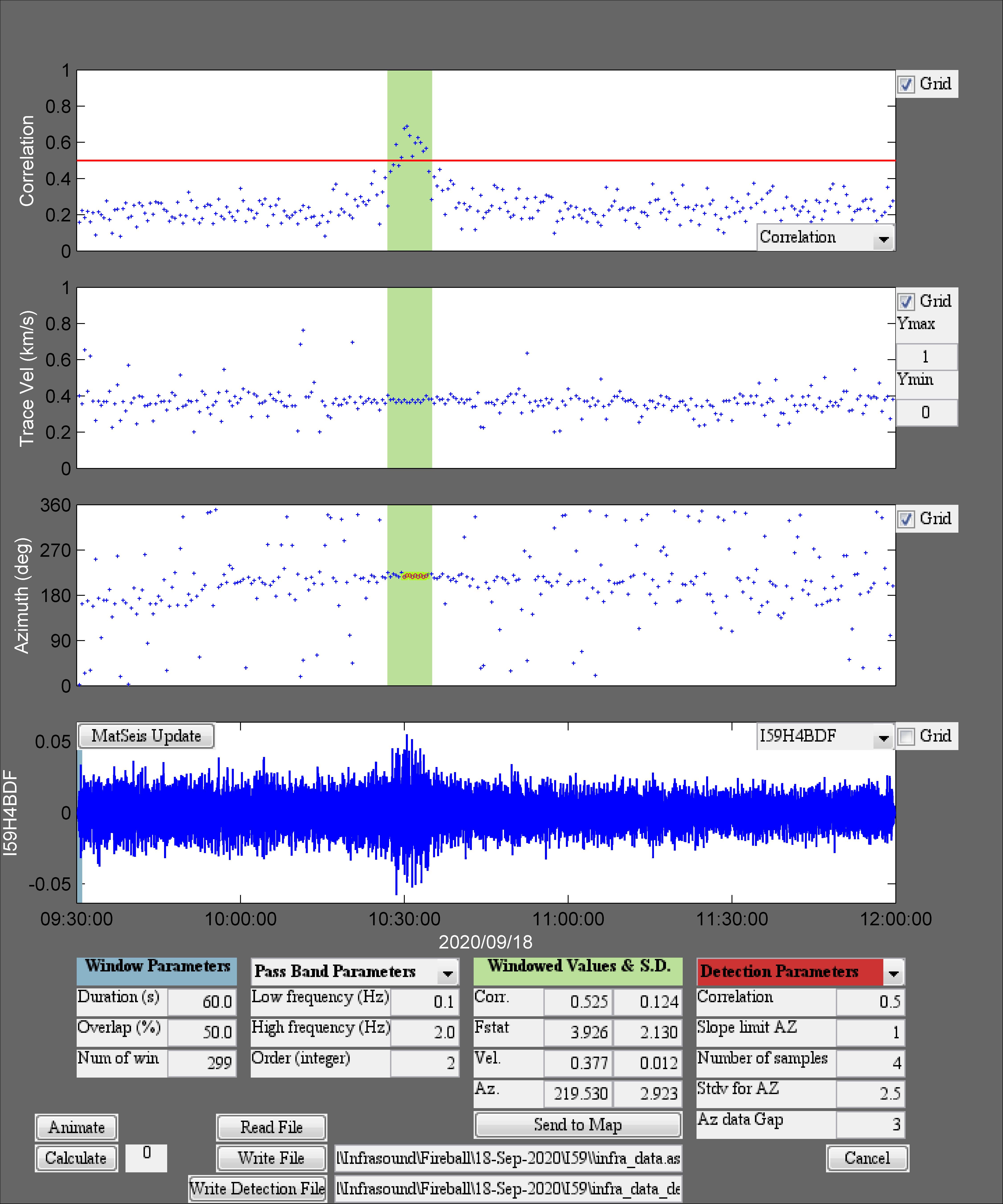}
\caption{Infrasound detected at the four element IS59 array in Hawaii from the USG 20200918 bolide. The bolide signal is centred around 10:30 UTC. Shown is the bandpassed waveform between 0.1 - 2 Hz (bottom plot) in units of Pascals for element 4 of the array. Here we have used 60 sec windows with 50\% overlap and found the best beam correlation (top plot), array trace velocity (second from top) and apparent backazimuth (third plot from top).  The main airwave arrival is highlighted by the vertical green bar. From I59 the great circle back azimuth to the USG fireball location is 218$^\circ$, while the average observed back azimuth of the signal across all frequencies centred at 10:30 UT is 219.5$\pm$2.9$^\circ$.} 
\label{infra}
\end{figure}

Infrasound (low frequency sound) from the bolide was detected at the IS59US array in Hawaii at a range of 2400 km beginning just after ~10:20 UT as shown in Fig \ref{infra}. Using the approach in \citet{Ens2012} all four array elements were stacked from the apparent direction of the airwave arrival to find a period at maximum amplitude from zero-crossings (cf. \citet{ReVelle1997} of 4.1 $\pm$ 0.3~s. Using the Progressive Multi-Channel Correlator approach of \citet{Cansi1995} we derive a period of 3.7$\pm$0.6~s, in agreement within uncertainty. The beamformed max pressure signal amplitude is 0.05 Pa using a bandpass from 0.1-2 Hz. 

Taking this period at maximum amplitude and the single station empirical yield-period relationship for bolides from \citet{Ens2012} we estimate a nominal source yield of 0.6  kt TNT equivalent, but with an uncertainty range of [0.2, 0.7]~kt. Using the  bolide-amplitude yield relations in \citet{Ens2012} and a computed stratospheric wind index (SCI) of 6.7~m~s$^{-1}$ using the ECMWF atmosphere model from source to IS59US, we derive a wind-corrected yield of 0.35~kt. 

Taken together, the available infrasound records suggest an energy in the 0.3~-~0.6~kt range as most probable. 

The published USG data (see Table \ref{tab:elements}) include the event time, location, velocity, height of peak brightness and total radiated energy together with a lightcurve of radiant intensity as a function of time. The latter can be used to estimate the total impact energy based on cross calibrations for events simultaneously detected by other techniques (infrasound, meteorite recoveries, ground based optical measurements) \citep{Brown2002, Edwards2006}. Among these USG sensor measured metrics, the radiant and to a lesser extent speed have been shown to be the least accurate quantities through comparison with high precision ground-based trajectory measurements, while the height of peak brightness has tended to be more accurate, typical errors being ~3 km \citep{Brown2015, Borovicka2015, Devillepoix2019}. The fireball location and energy are typically most secure, the former being more limited by the published precision. 

For the USG energy, we can compare to the GLM-derived energy. Here we follow the procedure outlined in \citet{Jenniskens2018} where the total energy in the GLM bolide lightcurve is summed, range corrected and a 6000~K blackbody spectrum assumed. Under these assumptions, a total optical energy of 5$\times$10$^{10}$ J is found from the GLM lightcurve, in good agreement with the 5.6$\times$10$^{10}$ J we derive from integration of the original lightcurve from USG sensors. From the lightcurve maximum radiant intensity, the peak bolometric magnitude of the fireball was -19. 

The resulting total radiated energy may then be converted to a total energy using an estimate for the luminous efficiency ($\eta$). The standard approach is to use the averaged relation from \citet{Brown2002} which yields an energy of 0.18 kt and $\eta$ = 0.07. This is 2-3 times lower than the infrasound yield estimates using either amplitude or period. As the generic relation derived by \citet{Brown2002} was based on a collection of bolides having an average speed near 18 km s$^{-1}$, substantially above the 12 km s$^{-1}$ for the current case, it is probable the real luminous efficiency is lower than the average value given in \citet{Brown2002}.

\citet{Borovicka2020c} has proposed the most robust fireball luminous efficiency relation to date, using the velocity dependence proposed by \citet{ReVelle2001} and validated through modelling of meteorite-producing fireballs where material is recovered on the ground. Using that formalism, we expect a factor of $\approx$2 higher value for fireball luminous efficiency at 18 km s$^{-1}$ compared to 12 km s$^{-1}$. Adopting this change for $\eta$  produces an energy estimate of 0.36 kt TNT and agreement between both optical estimates and the infrasound energy estimates within uncertainty. We suggest that the most likely energy for the event was close to 0.4 kt TNT on this basis.   

\begin{figure}
\plotone{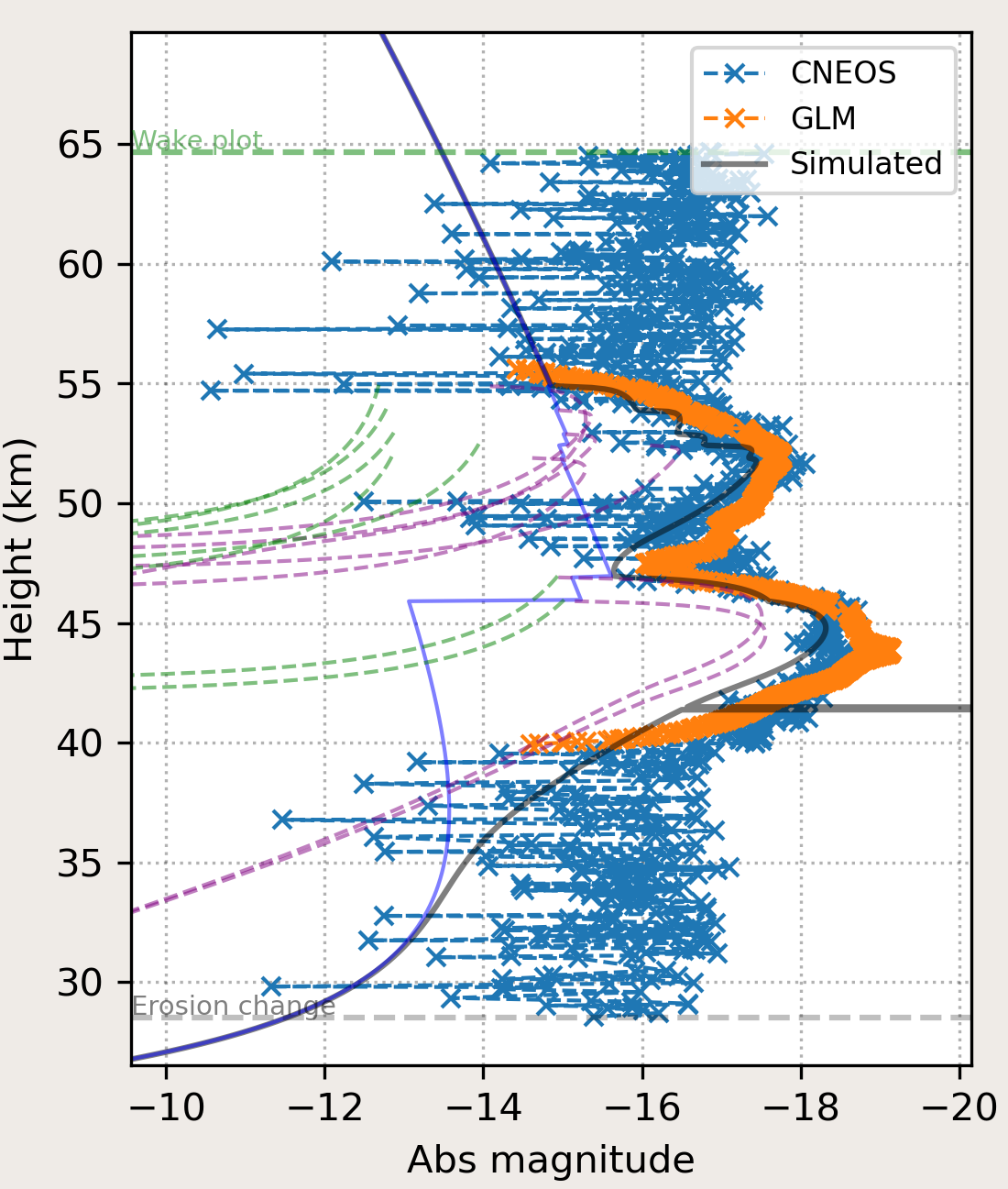}
\caption{The observed fireball light curve as a function of height as measured by US government sensors (blue) and by the GLM instrument (orange). The GLM lightcurve has been computed assuming a 6000~K blackbody following the procedure of \citet{Jenniskens2018}. Also shown is the lightcurve model fit (see text for details). Individual fragment lightcurves are shown as green hatched lines, associated dust released as erosion as purple hatched lines and the main fragment lightcurve is shown as a blue solid line. }
\label{fig:lc-fit}
\end{figure}

\subsection{Physical characteristics} \label{sec:physical}
\begin{figure}
\plotone{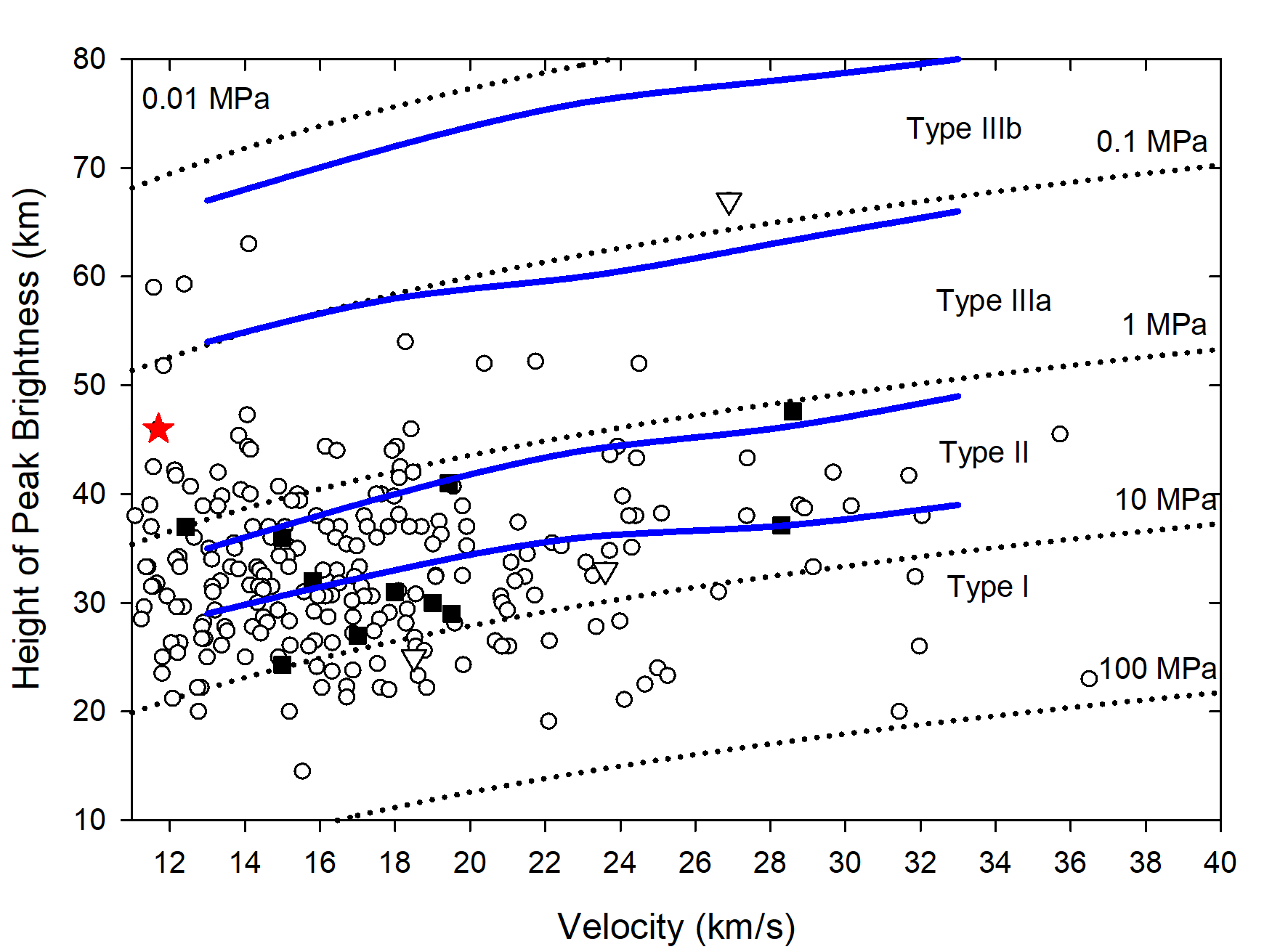}
\caption{Height at maximum luminosity as a function of entry velocity for meter-sized impactors. The USG 20200918 fireball is shown as a red star. The dotted lines are constant values of dynamic pressure and provide a proxy for strength. Black squares are meter-sized meteorite dropping fireballs, while inverted triangles represent ground-based fireball network observations of meter-sized objects \citep{Brown2015}. Open circles are from the CNEOS-JPL fireball database. See text for description of blue lines associated with various fireball types using model interpretation. }
\label{fig:fireballs}
\end{figure}

From these data alone we can place this event in some physical context. Figure \ref{fig:fireballs} shows the height at peak brightness for all meter-sized Earth impactors published to date. The majority of these are from the CNEOS-JPL site, but half a dozen (black squares) are the height of peak brightness from fireballs which produced meteorites \citep{Brown2015, Borovicka2021a, Jenniskens2021}. 

In general, a fireball's peak brightness is reached well after the first fragmentation episode, implying that the dynamic pressure at the height of peak brightness is an upper limit to the global strength of the object. Most meteorite producing fireballs have global compressive strengths of order 1 MPa \citep{Popova2011, Brown2015}, but reach the point of peak brightness when dynamic pressure is several times this value. 

As every fireball fragments differently depending on its collisonal history, entry models can at best capture only a rough correspondence of expected height of peak brightness as a function of global compressive strength, under the assumption of some standard fragmentation model. Here we use the triggered progressive fragmentation model (TPFM) described by \citet{Revelle2005}. \citet{Ceplecha1976} showed that fireballs can be approximately divided into four groups (I, II, IIIa, IIIb) progressing from strongest to weakest. These have been tentatively associated with stronger (ordinary chondrite) meteorites, carbonaceous chondrites and weak (and weaker) cometary-like material respectively \citep{Ceplecha1998}. Using this categorization, \citet{ReVelle2002a} adopted compressive strengths within the TPFM framework at which these various classes on average first fragment in the atmosphere, progressing from 0.7 MPa for Type I fireballs, to 0.2 MPa for Type II, 0.01 MPa for Type IIIa and 0.001 MPa for Type IIIb.

Using these values for strength and a representative energy of 0.4 kt for a typical impactor from the JPL CNEOS dataset \citep{Brown2015} the TPFM estimate for the expected height of peak brightness by fireball type as a function of entry speed is shown in Figure \ref{fig:fireballs} (blue lines).  

Notably, most of the fireball producing meteorites fall in either the Type I or Type II category as expected. Among the best comparator event for USG 20200918 is the Almahata-Sitta meteorite fall \citep{Jenniskens2009}. This meteorite was produced from the impact of 2008 TC3 which entered at a very similar speed as USG 20200918. It was the first preatmospheric asteroid impactor imaged telescopically. It entered at a much shallower entry angle than USG 20200918 and was several times more massive \citep{Borovicka2009a} and produced an extended series of flares near 40 km. The 2008 TC3 fireball fragmented under just a few tenths of MPa pressure, providing clear evidence of the very weak nature of the Almahata-Sitta meteoroid, which subsequent analyses using multiple techniques established as a probable rubble-pile assemblage with roughly 50\% porosity \citep{Welten2010, Borovicka2009a, Kohout2011} and an estimated bulk density of 1660 kgm$^{3}$\citep{Welten2010}.  

Given the higher height of peak brightness for USG 20200918 and even lower speed than Almahata Sitta, it is probable that it was similarly weak. Presuming its earliest fragmentation occurred 10-20 km above its height of peak brightness as suggested by the TPFM modelling (for comparison Almahata Sitta showed evidence for fragmentation as high as 53 km \citep{Borovicka2009a}, some 16 km above its point of peak brightness) suggests a global strength $\leq$0.1 MPa. This would place the meteoroid in the C or possibly D low strength meteoroid categories of \cite{Borovicka2020c} and be consistent with either a rubble-pile assemblage, collisionally re-assembled material or have very high microporosity. 

The other three meteorite producing fireballs from meter-sized or larger objects which show height of peak brightness at comparable dynamic pressures to USG 20200918 near the 1 MPa line are Kosice (H5 ordinary chondrite) \citep{Borovicka2013a}, Flensberg (C1-ungrouped) \citep{Borovicka2021a}, and Sutter's Mill (CM2) \citep{Jenniskens2012}. The weakness of the two carbonaceous chondrites is as expected given their material properties. Kosice was shown to be an unusually weak meteoroid for an H5 chondrite fall, fragmenting initially under a dynamic pressure of only 0.1 MPa \citep{Borovicka2013a}. 

To model the fireball in more detail we attempt to match the GLM and USG lightcurves using the semi-empirical fireball ablation model of \citet{Borovicka2013a, Borovicka2020c}. We fix the initial mass at 23000 kg (corresponding to our preferred energy of 0.4~kt) and an initial speed of 12.7~km~s$^{-1}$ (see next section). We assumed $\Gamma$A = 1.21, an ablation coefficient of 0.08 s$^2$~km$^{-2}$ and  then generate a synthetic lightcurve using the luminous efficiency model of \citet{Borovicka2020c}. Note that the resulting fit is not unique but representative of a family of possible fragmentation solutions. This is particularly the case as we have no detailed dynamics to simultaneously fit.

Our resulting fit is shown in Figure \ref{fig:lc-fit}. The main feature of the lightcurve is the two prominent maxima separated by almost 10 km in height. We find that the first maximum can be reproduced assuming release of half a dozen fragments each of order a tonne in mass which erode (release grains) rapidly in the early flare. These initial boulders are released under extremely low dynamical pressure of 0.1 MPa. 

The second flare can be matched assuming release at dynamical pressures of 0.3 MPa of another four multi-tonne masses which erode even faster than the first set. 

These fragmentation episodes are most consistent with the weaker strength classes defined by \citet{Borovicka2020c}, namely class C (or the very weak end of class B) objects. In the Borovicka classification, class C objects are interpreted as being loosely bound material which has reassembled following catastrophic collisions. These could resemble weak macroscopic fragments loosely bound in a rubble pile formation but lacking significant interstitial dust. 

Thus the overall picture which emerges for USG 20200918 is of a weakly bound object of 3 m diameter on an Earth-like orbit. We note that among the USG reported meter-sized objects, there is a peculiar population of four other very weak objects which impact Earth at low speed (and have Earth-like orbits) which occur just above USG 20200918 in Figure \ref{fig:fireballs}. Why these do not occur at higher speeds is unclear. We do note that in order to  reconcile the 3 m diameter with the magnitude H=32.5 determined from the light curve in Section \ref{sec:lightcurve} a an extremely low albedo of approximately 0.02 must be used.  We discuss this further below.  

\section{Telescopic Light curve of the ATLAS detection}\label{sec:lightcurve}

\begin{figure}
\plotone{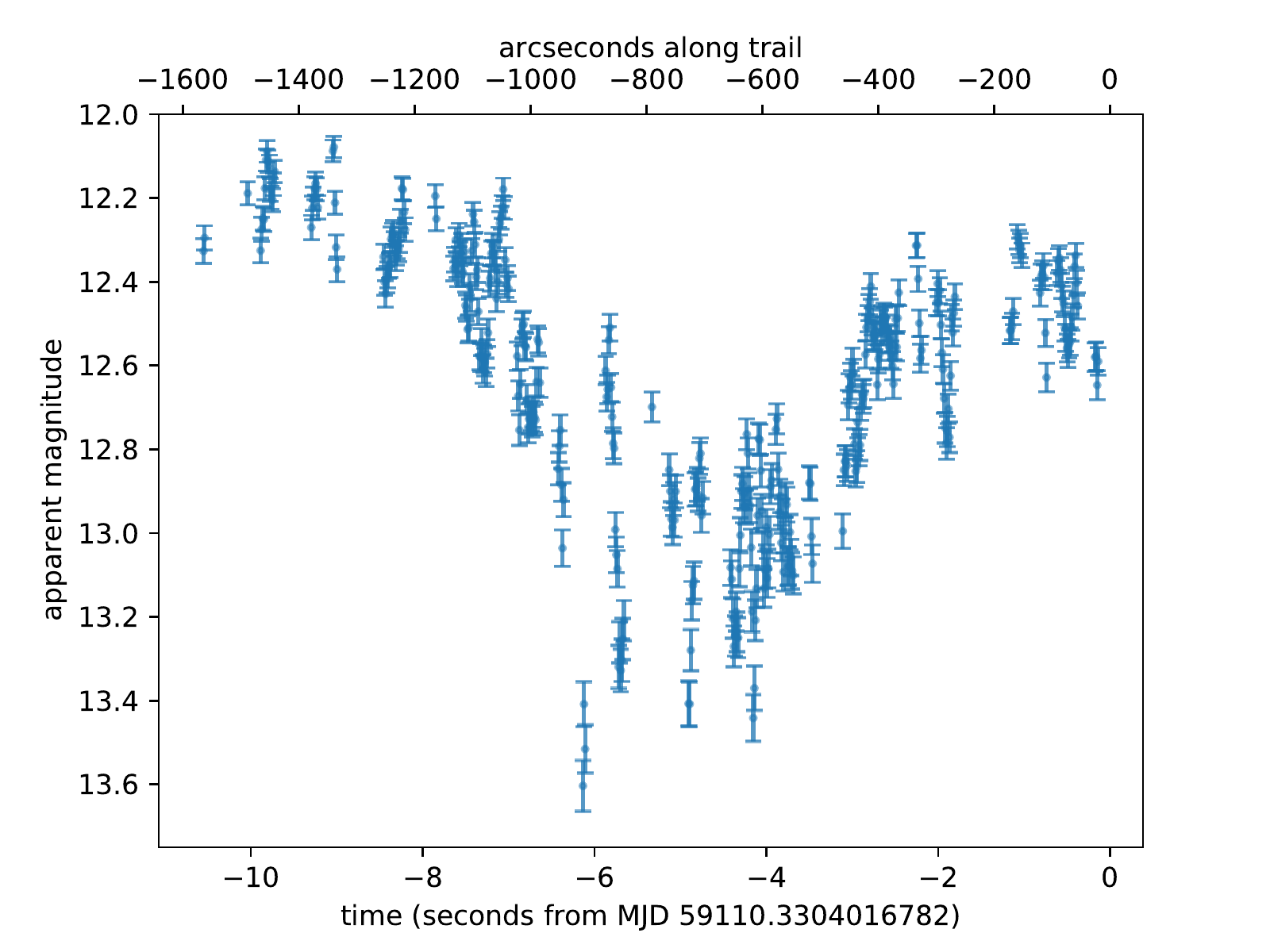}
\caption{Apparent c-band (cyan) magnitude along the ATLAS trail. Time and trail length are measured from shutter closure, assuming the rate of motion calculated in this work. Regions of the trail less than 8 pixels from the center of a star have been masked out.}
\label{fig:lightcurve}
\end{figure}

The light curve seen by ATLAS is shown in Figure~\ref{fig:lightcurve}. It is extracted from the difference image computed by subtracting from the original image a low-noise stack of prior images convolved to the PSF of the original image \citep{2018AJ....156..241H}.  With this method of image subtraction the original image's photometric calibration and noise properties are preserved. Though stars are largely removed by the differencing process, some artifacts remain. As a result, all portions of the trail which are less than 8 pixels from the centroid of a star are masked out. The remaining portion of the trail (which is 1582 arcseconds long, corresponding to 10.7 seconds of time assuming our calculated average rate of motion of 148 arcseconds per second) is sampled with 352 overlapping square apertures 6 pixels on a side, using astropy's photutils package \citep{price2018astropy}.

The light curve demonstrates the smoothly varying bright-dim-bright pattern which is visible to the eye in Figure~\ref{fig:actualimage}. The time from peak to peak is 8 seconds, yielding a rotation period of roughly 16 seconds assuming a simple double-peaked light curve. The amplitude is about one magnitude, implying an axis ratio of $\sqrt{2.5}\approx 1.6$. The object was seen at an apparent magnitude that is one to two magnitudes fainter than the predicted value of 11.5.  No accompanying streaks indicative of a companion object were seen during visual inspection. 

Could the object be a meteor?  First, the variation in lightcurve across the image and abrupt end are not consistent with a meteor trail \citep{Beech2006}. The abrupt end, in particular would mean the exposure had to stop during the luminous flight which is highly unlikely given typical meteor durations at such magnitudes are of order a few tenths of a second \citep{Fleming1993}. Secondly, if the object were an actual meteor it would have to be much  closer (range of order 100 km), and so it would be out of focus \citep{2004M&PS...39..609J}. 

\begin{figure}
\plotone{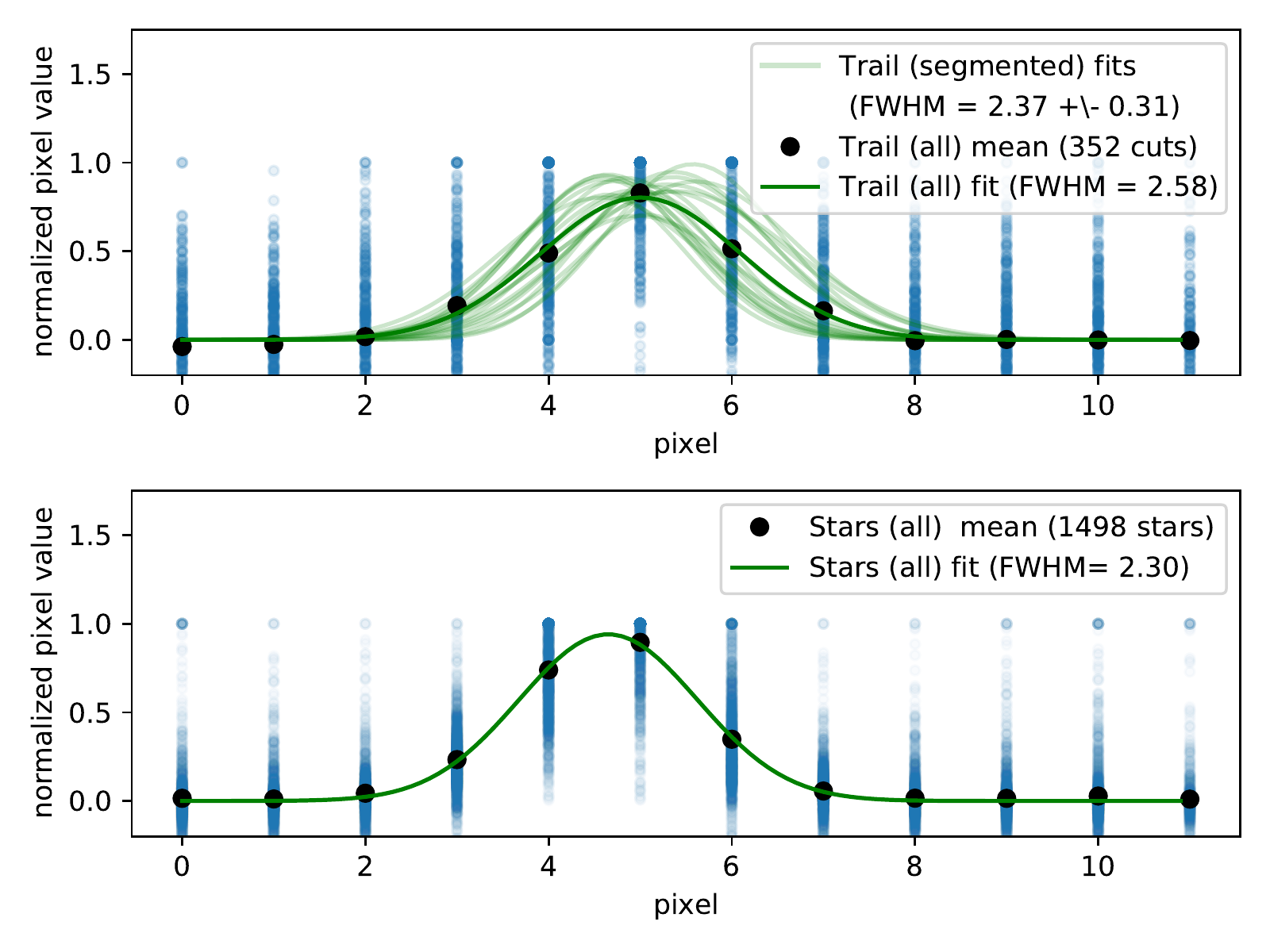}
\caption{The cross-sectional width of the ATLAS image trail. In the upper panel, the best-fit Gaussian cross-section of the entire trail with respect to a hypothetical straight line between the trail's start and end is shown in dark green. This fit has a FWHM of 2.58 pixels. However, the trail has a slight curvature which broadens this distribution. When the trail is broken up into short segments and fit individually (light green lines), these show a smaller FWHM (2.37 $\pm$ 0.31) which is consistent with the stars in the image. This result indicates that the trail is not close enough to the telescope to undergo any de-focusing. In the lower panel, cross-sections of nearby stars,  sliced along the same direction used to construct the trail cross-sections, are shown for comparison.}
\label{fig:crosssections}
\end{figure}

To examine whether there is any sign of the trail being out of focus relative to the stars, the cross-sectional profiles of the trail and of nearby stars was extracted as shown in Figure~\ref{fig:crosssections}. The average profile for the stars is constructed from 1498 stars near the trail and sliced along the same direction as cross-sections of the trails. The stars have a FWHM of 2.30 pixels under a 1D Gaussian fit. This is consistent with the ATLAS image's overall stellar FWHM of 2.42 pixels, computed automatically by their software. The profile for the trail is constructed from cross-sections taken at the locations of the apertures used to extract the light curve of Figure~\ref{fig:lightcurve}. The trail displays a slight (approximately one pixel) curvature from one end to the other, probably due to optical effects.  As a result, stacking all 352 cross-sections and then fitting a Gaussian profile slightly overestimates the trail's width, coming in at 2.58 pixels FWHM. To better estimate the true width of the trail, we split it into 20 segments, each consisting 17 to 18 adjacent cross-sections, chosen empirically to provide reasonable time resolution and sufficient signal-to-noise. The mean FWHM of the trail obtained in this manner is $2.37 \pm 0.31$ pixels, effectively the same as that of the stars. We conclude that there is no evidence that the trail is close enough to the telescope to undergo de-focusing, and hence cannot be a meteor or any in-atmosphere object.

\section{Combining the \textbf{USG} and ATLAS Astrometric data}\label{sec:combining}

The USG fireball detection and the ATLAS image streak provide two independent constraints on the location of the object in question. In this section we ask: are they dynamically consistent with each other?  We will see that the answer is yes, solidifying the linkage of these two objects as well as providing some refinement of the USG velocity measurement.

To investigate this question, dynamical simulations of the USG bolide were performed to determine whether or not its motion could bring it to the correct location on the ATLAS image.  The simulations used the RADAU \citep{1985ASSL..115..185E} integrator to model the object along with the Earth, Moon, Sun and other planets (as derived from the DE405 ephemeris \citep{sta98}) backwards to the time of the ATLAS image.

The Bayesian fitting package Multinest \citep{multinest2011} was used (through its Python interface PyMultinest \citep{pymultinest2016}) to determine what values of the fireball position and velocity components produce a best-match to the streak seen on the ATLAS image.

The best fit was determined by minimizing a $\chi^2$ function to four observables; namely, the RA and Dec of the end of the streak on the ATLAS image (304.0814$\degree$ and +7.1452$\degree$ respectively, each $\pm$ one pixel = 1.86 arcsec $\approx 5 \times 10^{-4}\deg$), the on-image angle of the streak (122.25 $\pm 0.3\degree$ clockwise from the lower edge of the image) and the apparent magnitude ($12.5 \pm 0.2$)

Seven parameters were fit, six with Gaussian priors centred on the nominal USG position/velocity values listed in Table~\ref{tab:cneos}. Since the USG reports do not provide uncertainties, our priors assumed $\pm 1$~km s$^{-1}$ for each of the velocity components, $\pm 0.25\degree$ for the latitude and longitude, and $\pm 10$ km for the altitude of the bolide.  The seventh fitted parameter, the absolute magnitude, is not reported so we adopted a uniform prior for $H$ running from 31 to 34. 

\begin{figure}
\plotone{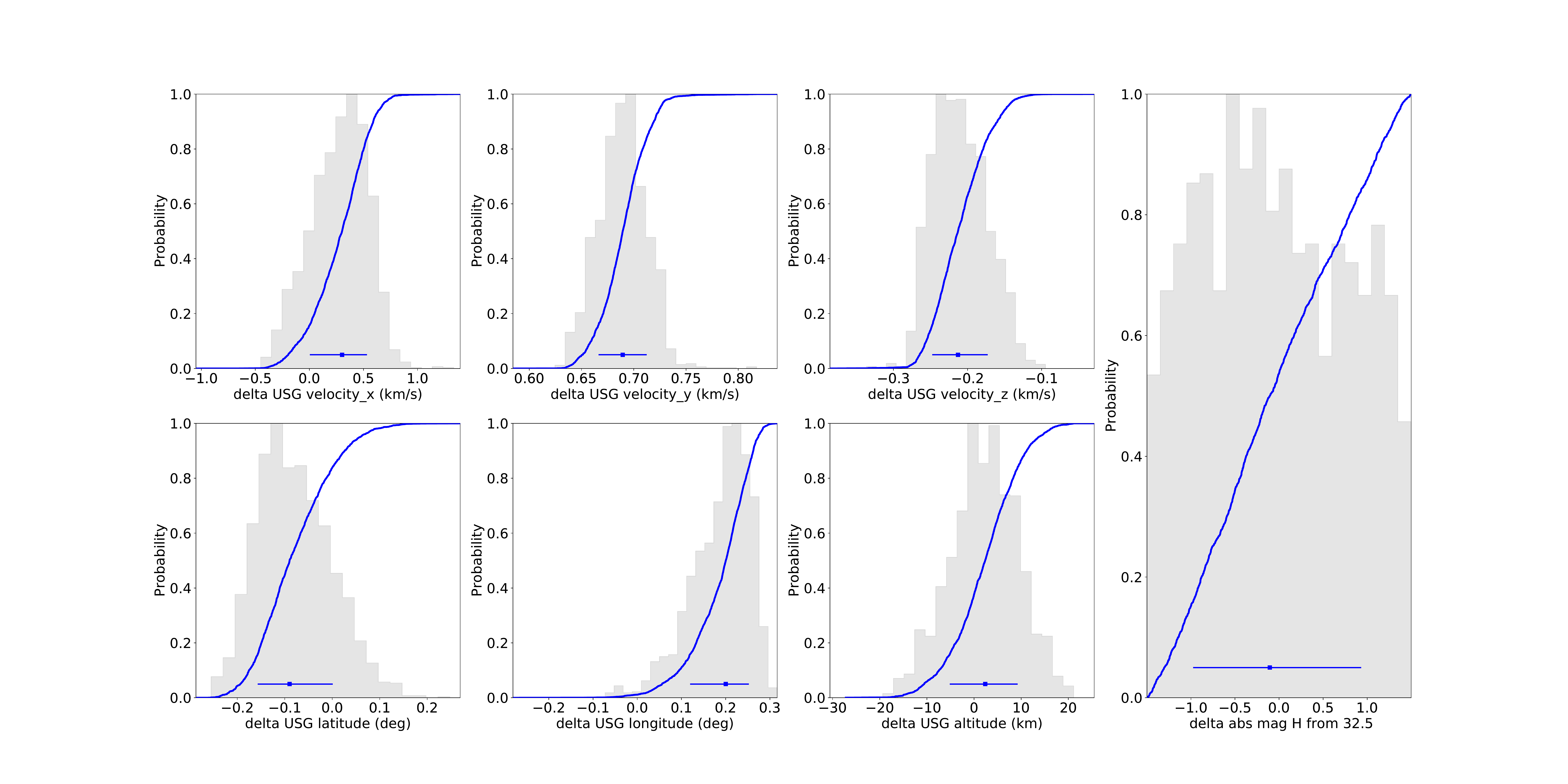}
\caption{Marginal probability distributions (in grey) from one of the parameter fitting runs. The blue points with horizontal uncertainty bars show the median values with their one-sigma uncertainties. The fitted parameters are displayed as differences from the nominal USG values and from a nominal absolute magnitude $H$ of 32.5.  The blue curves shown the cumulative probability distribution.}
\label{fig:statranging}
\end{figure}

The fitting process was repeated 10 times to ensure consistency. There were no statistically significant differences between the fits, so the result of a typical run are presented here. The marginal probability distributions of the fitting are displayed in Figure~\ref{fig:statranging}. All the fits are consistent with the USG values, with the exception of a fraction of a degree shift in lat/lon and a preference for a slightly higher velocity. The overall best-fit (lowest $\chi^2$) values are shown in Table \ref{tab:cneos} and will be adopted here as our best estimate of the object's true contact state. This gives a refined velocity vector with a total speed of 12.7 km s$^{-1}$ versus the nominal USG value of 11.7 km s$^{-1}$.  The overall parameter fit matches the termination end of the streak within an arc-minute, the slope angle to within 0.168 of a degree, and the magnitude to within 0.14 magnitudes.” Back integrating an uncertainty cloud generated from the best fit values results in the orbit elements listed in Table \ref{tab:elements} and the modified orbit shown in Figure \ref{fig:orbitdiagram}.

From this result,  we can confidently conclude that the USG and ATLAS objects are completely consistent with each other offset by a small velocity difference that is well within previously reported USG speed uncertainties \citep{Devillepoix2019}. In fact, the best-fit values effectively tell us what the inclusion of the ATLAS image as a constraint does to the orbital solution. By folding that information in, we can see that the object's arrival velocity was actually slightly higher than that derived by consideration of the USG data alone. This provides additional confidence that the object was not in a geocentric orbit, pushing it clearly to an arrival from heliocentric orbit. 

The absolute magnitude of the object is only loosely constrained, but the best-fit value of $H=32.5$ corresponds to a 0.8, 1.8 and 3.0 m diameter for albedos of 0.25, 0.05 and 0.02 respectively. This makes it one of the smallest asteroids ever observed in space, as will be discussed in the next section.

\section{On-image Object Origin} \label{sec:origin}

\begin{figure}
\plotone{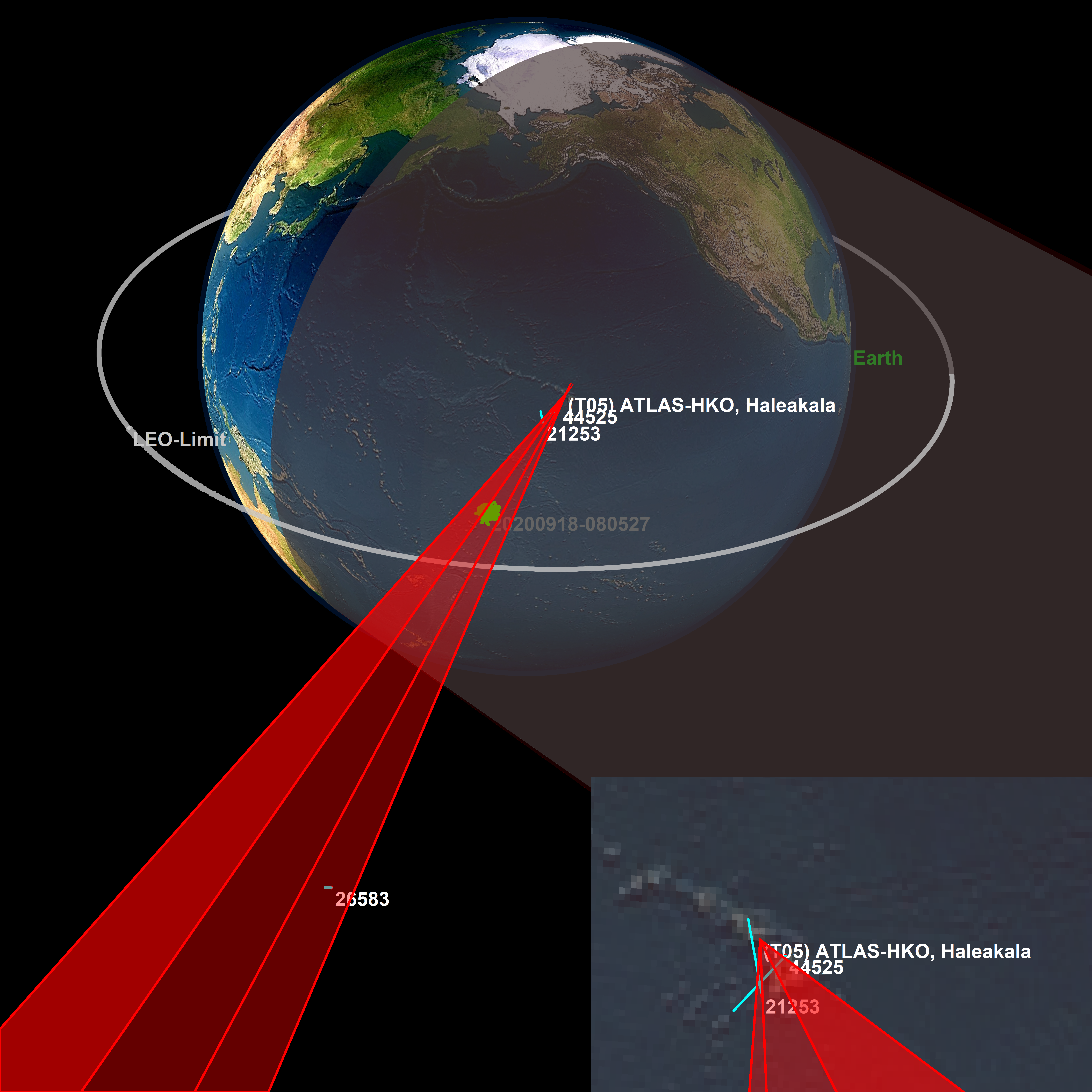}
\caption{An illustration of the Earth's shadow alignment at the time of image capture. The umbral shadow is shown in dark gray.  The penumbral shadow is imperceptibly larger than the umbral at such close proximity to the Earth.  The gray circle is the upper "limit" of LEO orbits at $H_{LEO}=2000$ km.  The red wedge is the the volume of space covered by the image.  The irregular green area is the predicted uncertainty of the object. The numbered blue lines in both the main image and inset represent the motions of all tracked artificial objects in the field of view of the image, with their NORAD TLE numbers.   Within the field of the image, any object inside the LEO-limit, and substantially further, is in the Earth's shadow and would not be visible in the image.  Only three tracked objects are within the image field, and only one of them (26583) is outside the Earth's shadow and is visible.  This object is evident in the ATLAS image reproduced in Figure \ref{fig:actualimage}.}
\label{fig:shadowdiagram}
\end{figure}

\begin{figure}
\plotone{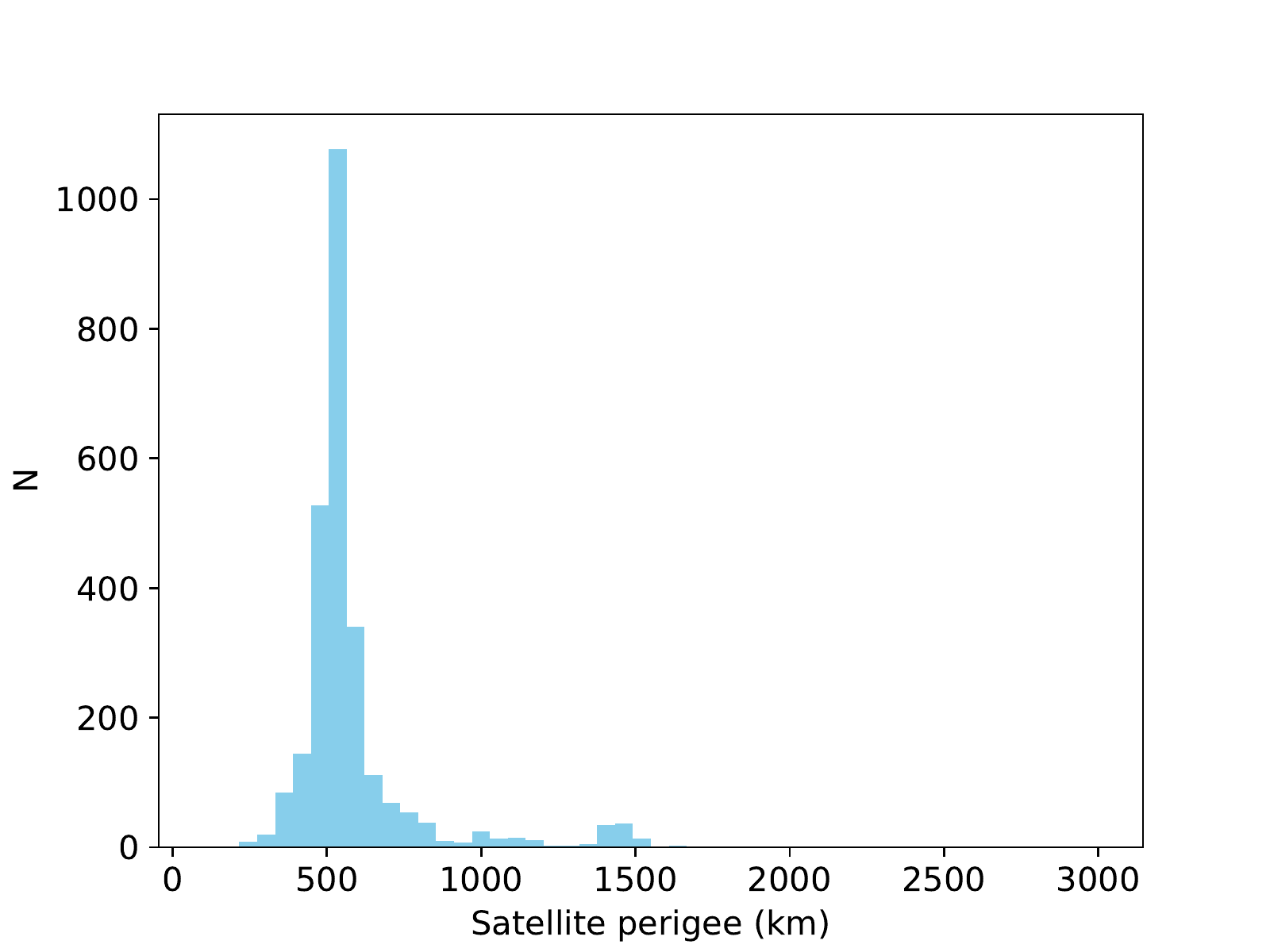}
\caption{A histogram of Earth satellite perigee values, from the Union of Concerned Scientists Database retrieved May 13 2021.} 
\label{fig:satelliteorbits}
\end{figure}

The USG fireball and the ATLAS object could either be the same object, or two different ones. Since Occam's razor suggests they are the same object, we will discuss this possibility first, and then return to the less likely case. If the USG and ATLAS detections are the same object, then it is almost certainly natural.

First, the USG entry orbit is not bound to Earth. Though there is man-made material in heliocentric orbit, the bulk of it is geocentric orbit and the coincidental re-entry of a substantial man-made object from heliocentric orbit is extremely unlikely. Moreover, the high inclination of the USG orbit is not consistent with a space-age launch into a heliocentric orbit as it far exceeds typical escape speeds for interplanetary missions. 

The impact speed of 11.7 km s$^{-1}$ puts it just above the unbound-from-Earth threshold at 11 km s$^{-1}$. Could measurement uncertainty be masking what was actually a bound orbit?  To obtain a rough estimate of the possible velocity error (the \textbf{USG reports}  no uncertainties), we note that ATLAS was taking 30 second exposures; and that 30 seconds earlier or later and the object would have been off the frame. ATLAS images are about 5 deg wide, which at 6000 km corresponds to a distance of $6000$~km~$\sin 5\degree$ = 500 km. Assuming the worst case of tangential motion, a 500 km difference over the less than 10 minute ($\sim 600 $ sec) interval between the ATLAS image and the USG detection require a velocity error of about 1 km s$^{-1}$. So we cannot strictly exclude a velocity of 10.7 km s$^{-1}$, that is, on a bound ---but only just-- orbit from the ATLAS data. Any normal component to the object velocity would mean an even higher speed, further increasing the likeliness of the object having an unbound orbit.

If the object was artificial, it must have been large. The initial estimated mass of 8600 kg derived from the USG energy of 0.14 kt is inconsistent with a small highly reflective rocket body or satellite. For comparison the total mass of the Soyuz MS spacecraft is 7080~kg; its descent module, 2950~kg\footnote{\url{https://en.wikipedia.org/wiki/Soyuz\_(spacecraft)} retrieved 2021 May 20}. The unexpected arrival of such a large man-made object is extremely unlikely. 

The radiant azimuth of the USG detection at 13 degrees from north would put it on a hypothetical geocentric orbit near 103 degrees inclination to Earth's equator if its speed were low enough to be bound. This is within the range of inclinations typical of sun-synchronous orbits: 103 degrees corresponds to a sun-synchronous orbit with a 2 hour period, orbiting at an altitude of 1681 km\footnote{\url{https://en.wikipedia.org/wiki/Sun-synchronous\_orbit} retrieved 2021 May 20}. However all such large satellites are catalogued and no predicted re-entries occurred near the impact time. Moreover any satellite decaying from LEO would have a much lower re-entry speed, nearer 8 km s$^{-1}$ and outside our bounding range. Thus if the ATLAS and USG detections are the same object, they are very difficult to reconcile with an artificial object.

If the USG and ATLAS detections were separate objects, could the ATLAS telescopic image be a coincidentally-located artificial object? If that were the case, the USG velocity must have been off by $>1$ km s$^{-1}$ (for the meteoroid to be off the ATLAS frame), and a different object travelling in the 1) right direction at the 2) right brightness 3) at the right time and at least approximately 4) at the right speed must have been in-frame. Fortunately all objects $>10$~cm are known, tracked and catalogued in LEO \citep{kessler2011} with catalogues publicly available at Space-Track.org\footnote{\url{https://www.space-track.org/}}. 

The speed is less well constrained because the trail starts (or ends) outside the image. But the object in the ATLAS image must travel at least 1579 arcsec (the length of the trail on image) in 30 seconds, or at least 53 arcsec per second. An object in LEO (which might be travelling 200 arcsec sec$^{-1}$ or more)  could make a comparable streak length, and at an apparent magnitude of 11.4, a satellite at 3000 km would be of order 1~m$^2$ cross-sectional area. The inclination is reasonable for a man-made object as well: if we assume the motion seen in the image is a circular orbit in LEO, then its orbital inclination is roughly the angle of the trail with the $x$-axis, that is 58 degrees (or 180-58=122 degrees depending on which direction it is moving). The former value is near the heavily populated region of LEO containing non-polar inclined orbits (peaking near 53 degrees, see Figure~\ref{fig:satelliteorbits}), so there are certainly satellites in this region.

Fortunately the geometry allows us to exclude objects in LEO because they would be in the Earth's shadow. The Sun had set at the ATLAS site at the time in question, and sunlight was still illuminating objects only at altitudes greater than 3400 km above sea-level. So all of LEO would have been in shadow while the meteoroid ---at an altitude of 6000 km--- would have been in full sunlight at the same time. Figure~\ref{fig:shadowdiagram} illustrates the geometry during the ATLAS observation. If the object were a satellite, it would have to be in the much more sparsely populated region above 3400 km. Though the possibility of an artificial object cannot be positively excluded to this point in our analysis, we can conclude that the ATLAS and USG detections have a very high probability of being the same natural object.

To fully eliminate the possibility of the ATLAS detection being that of a man-made object, we downloaded the full catalogue of element sets for all tracked orbiting objects from Space-Track.org\footnote{\url{https://www.space-track.org/}}.  We use the TLE (3 Line Element) formatted file\footnote{\url{https://www.space-track.org/basicspacedata/query/class/gp/EPOCH/\%3Enow-30/orderby/NORAD\_CAT\_ID,EPOCH/format/3le}}, although we have used the older NORAD TLE (two line element) terminology to refer to the catalogued objects.  The 21944 catalogued TLEs were used to plot the object trajectories with respect to the ATLAS image.  Three objects were found to intersect the image (see Figures \ref{fig:searchimage} and \ref{fig:satelliteorbits}): object 44525 passed though the field of view, while object 21253 and object 26583's paths begin in the image and terminates outside.  As can be seen in Figure \ref{fig:satelliteorbits} object 44525 and object 21253 are in LEO and therefore in the Earth's shadow.  They could not be (and are not)  visible in the image.  Object 26583 is well outside the Earth's shadow and is visible in the ATLAS image at the correct location, distinctly different from the candidate image streak. 

This exercise highlights the challenge of discriminating a telescopic pre-impact detection of a meter-sized fireball --- most of which have coarse state vector accuracy from in-atmosphere measurements--- from satellites. The proliferation of low Earth orbit satellites will make future associations harder. 

Another consideration is the possibility of the image streak being that of an in-atmosphere natural or man-made object such as a meteor, reentry debris, or airplane.  Section \ref{sec:lightcurve} describes a detailed analysis of the lightcurve of the ATLAS detection.  There is no evidence of de-focusing expected for such a near field object, so that possibility can be discounted.

Having argued the image streak is not due to an artificial or in-atmosphere object, we now address the possibility that the image streak was caused by another natural out-of-atmosphere object distinct from that reported by USG. To do this, we  will compute the number of objects expected to appear in the image at the correct magnitude and on-sky speed, and show that is it very low, making the coincidental appearance of another object very unlikely. 

\citet{Brown2002} provides a power law relating the diameter of an impacting object with the cumulative number of objects of that diameter or greater impacting the Earth in a year.  The number of impacts can be directly mapped to a near-Earth number-density of these objects by using a cylindrical volume of space determined from the cross sectional area of the Earth and the average speed of impactors (we use 20 km s$^{-1}$).  The sizes of objects to be considered are bound by two factors, the closest possible distance of the object based on the the object being outside the Earth's shadow, and the furthest possible distance of a solar system object that could result in a streak as long as that observed on the image.  Using standard relationships of absolute and visual magnitude, together with the object's phase angle, and relationships of asteroid diameter and absolute magnitude, we can estimate the diameter of both nearest and furthest possible objects that could result in the brightest (m = 12.1)  visual magnitude captured (Figure \ref{fig:lightcurve})\textbf{. }For these calculations we assume representative object albedos and slope parameters of 0.15. 

Extending the line of sight of the object streak outside the Earth's shadow results in a nearest object distance of approximately 3000 km with an object diameter of 0.73 m.  Assuming the worst case of an object moving at the geocentric solar escape velocity of 72 km s$^{-1}$, the furthest an object could be from Earth to produce the on-image streak subtending 0.442$\degree$ (Table \ref{tab:streak}) would be approximately 280000 km with an object diameter of 71 m.  A high estimate of the total number of observable m = 12.1 objects at any point in time is arrived at by summing the numbers of objects at incremental distances and magnitudes between the two limits, and dividing by the portion of the sky covered by the \textbf{ATLAS} image (5.5$\degree$$\times$5.5$\degree$ / 41000$\degree$$^{^2}$). At the near limit we arrive at 8 $\times$10$^{-8}$ objects expected on the image and 3 $\times$10$^{-7}$ at the furthest limit.  These numbers slowly increase with distance and required size, with the volume to distance of exponent of 3 being strongly offset by the 2.7 index of the \textbf{\cite{Brown2002}} power law and the log$_{10}$ relationship of magnitude and distance. Summing across regular intervals of distance we arrive at an approximate number of m = 12.1 or larger objects on any such \textbf{ATLAS} image of 3.5$\times$10$^{-6}$ objects.  This strongly suggests that the image streak is the same object as detected in the USG fireball observation and not an unassociated natural object. 

\section{Subsequent Image Search} \label{sec:Subsequent}
Subsequent to work performed to determine an object trajectory which best fit both the on image streak and the USG observations (see Section \ref{sec:combining}),  new image searches were performed based on this new trajectory.   These searches suggested two additional surveys may potentially have captured the object.

The first is an image quartet taken by the Catalina Sky Survey using the Mount Lemmon telescope\footnote{\url{https://catalina.lpl.arizona.edu/about/facilities}}.  The images G96\_20200918\_2B\_N27128\_01\_0001, \_0002, \_0003, and \_0004 were taken 2.5 to 2.0 hours prior to contact where the object would have been at an approximate visual magnitude 18.  Using 20,000 clones generated in the best-fit process, we calculate a 10$\%$ probability of the object being on the first image of the quartet, increasing to 20$\%$ on the fourth image.  We estimate a trailing loss of 4.5 to 5 magnitudes resulting in an effective visual magnitude of 22.5-23. With Mount Lemmon's limiting magnitude of 21.5 the object would likely not be visible in the images. Visual blinking of the image quartet using an automated blink and pan using the DS9 software \citep{2003ASPC..295..489J} did not yield any detection of a moving object.  Our software permits us to highlight the locations of known objects in DS9.  Objects significantly brighter were visible in the images, but objects of similar magnitude to our target meteoroid were not, consistent with the non-detection.   We also performed a visual scan of median subtracted images; again no object was found.  Finally, we performed a shift and stack process shifting the images by aligning the projected position of the object on each image, and stacking the results.  Stacking four images yields only a $\sqrt{4}/2.5 < 1$ magnitude increase in sensitivity.  Adjusting Mount Lemmon's limiting magnitude of $21.5$ to 22.5 due to stacking places the object at the limit of visibility.  Again, no object was detected in either the stacked original images or stacked median subtracted images.

Secondly, a recent implementation of a Zwicky Transient Facility (ZTF) \footnote{\url{https://www.ztf.caltech.edu/}} image catalogue {\footnote{\url{https://irsa.ipac.caltech.edu/frontpage/}}} import into our image database yielded an additional potential image capture. ZTF image ID taken 43 minutes before impact has a 99.9\% probability of containing the object.  If visible the object would appear as a 10 arcmin (566 arcsec) streak.  However, the object's visual magnitude of approximately 16.0 at this time combined with a trailing loss of nearly 7 magnitudes render the object much dimmer than ZTF's limiting magnitude of 20.5. Still, a visual scan was performed but no object was found.  

FROSTI image searches were performed against an ever-increasing set of current and historical image catalogues in an attempt to locate additional possible imagery: \textbf{ATLAS} Haleakal\a=a and Mauna Loa (and recently Altas Sutherland Observing Station and El Sauce), CFHT, Catalina (Mount Lemmon, Mount Bigelow, Catalina Follow-up, Kuiper, Bok, and Siding Springs), CNEOST, DeCAM, HST, LONEOS, WISE, Pan-STARRS 1 and 2, WISE and NEOWiSE, ZTF, and all contributing Observatories to the Minor Planet Centre Sky Coverage Pointing Data \footnote{\url{https://www.minorplanetcenter.net/iau/info/PointingData.html}}. No additional possible pre-impact images were found.

\section{Discussion} \label{sec:discussion}
  
The rotation period inferred for the object studied here is short but not unprecedented. \cite{Hatch2015} examined the published data for the 88 smallest known near-Earth asteroids (all with diameters less than 60 m) and found that the average rotation period of the sample is 40 min, with smaller asteroids rotating faster. Three asteroids in their sample were found to have rotation periods (retrieved from the Asteroid Light Curve Photometry Database (ALCDEF)\footnote{\url{www.alcdef.org}} \cite{ALCD2009}) below 60 s:  2010 WA (3 m diameter S-type) with a period of 31 s; 2010 JL88 (13 m diameter S-type) with a period of 25 s; and 2014 RC (13 m diameter S-Type) with a rotation period of 15.8 seconds. The first two of these asteroids have a quality rating in ALCDEF of $U=3$, which is the highest rating, while the third has no rating. We conclude that the rotation period of 16 seconds inferred here is inline with that of other known small asteroids.

\cite{Hatch2015} also collected data on the 92 smallest NEAs with known axis ratios; this sample includes diameters up to 300~m. They found that the mean and median axis ratios are 1.43 and 1.29, respectively, though axis ratios above 2 were not uncommon. We conclude that the axis ratio of 1.6 for this event as determined from the ATLAS light curve (Figure~\ref{fig:lightcurve}) is consistent with those of other small NEAs.

Assuming a rotation period of 16 sec and diameter of 3 m we can also work out the minimum binding strength of the meteoroid. Using the formula for the limiting tensile strength at the center of a rotating sphere able to just provide sufficient centripetal acceleration to keep the sphere intact from \citet{Kadish2005} for a 3 m diameter sphere with bulk density of 1600 kg~m$^{-3}$ we find a limiting strength of $<$1 kPa. This is well below the earliest fragmentation pressure and therefore physically plausible. 

The 3 m diameter determination from Section \ref{sec:fireball} and the best fit absolute magnitude H=32.5 from Section \ref{sec:combining} necessitate a very low albedo of 0.02 under the standard assumptions of asteroid photometry.  Is this reasonable? Only $10\%$ of NEA's observed by the NEOWISE have albedo less then 0.03, 5\% less than 0.02 \citep{2016AJ....152...79W}.  However, fully 25\% of NEOWISE-observed NEA's belong to an albedo population peaking at an albedo of 0.03.  Factors which could lead to higher visual magnitudes and therefore higher absolute magnitudes and lower albedo estimates include shape profiles and self-shadowing.  Assuming an oblate spheroid, the above estimated 1.6 axis ratio could easily result in a smaller face of the object reflecting sunlight to the observer than would be assumed for a spherical object.  This is of course an argument of possibility,  in that an object could equally have a larger face reflecting. If the object is more irregular in shape, as would be reasonable for a low-mass rubble-pile, then self-shadowing would begin to take effect, lowering the overall brightness of the object for an equivalent albedo. Self-shadowing is more influential at higher phase angles.  With the USG object at a phase angle at nearly $56\degree$ self-shadowing could very well be a factor in the larger magnitude estimation.

Given that the object arrived on the orbit listed in Table~\ref{tab:elements} we might ask about its likely escape region from the main belt. The orbit is certainly consistent with an evolved asteroidal orbit rather than a cometary one. The NEO model of \cite{Granvik2018} provides source region probabilities as a function of $a$, $e$, $i$ and absolute magnitude $H$. The $H$ of our object at about 33 is well below the range of $17 < H < 25$ considered by \cite{Granvik2018}, but if we examine the source region probabilities of their smallest model bin (central $H = 24.785$) we find the probability is 70\% for the main-belt $\nu_6$ resonance, 19\% from the Hungarias, 11\% from the 3:1 and other source regions negligible.

The low albedo and structural weakness of the object would be consistent with a C-complex NEA. Given the high probability of escape from the $\nu_6$ resonance it is interesting to note that \citet{Marsset2022} have found that the debiased fraction of C/D/P NEAs are almost 40\% of the total with present orbits associated with that source region. They also report a higher abundance of D-type NEOs than previously found from all escape regions, including the $\nu_6$. In this context, that the USG 20200918 fireball appears consistent with a C-complex meteoroid is not surprising. 

\begin{deluxetable*}{lrlrlrlr}
\tablenum{7}
\tablecaption{Nearest Preatmospheric Observations of Natural Objects\label{tab:nearObs}}
\tablehead{
Object & $d_{min}$& \multicolumn2c{Initial Observation} & \multicolumn2c{Last Observation} & \multicolumn2c{Closest Observation}\\
\multicolumn2r{(au)} & \multicolumn1c{Date/Time} & \multicolumn1c{$10^3$km} & \multicolumn1c{Date/Time} & \multicolumn1c{$10^3$km} & \multicolumn1c{Date/Time} & \multicolumn1c{$10^3$km}
}
\startdata
\multicolumn2l{Earth Impactors} \\
\hline
2008 TC3&&2008 Oct 6.27767&493&2008 Oct 7.07310&34&2008 Oct 7.07310&34\\[-5pt]
2014 AA&&2014 Jan 1.26257&417&2014 Jan 1.31081&396&2014 Jan 1.31081&396\\[-5pt]
2018 LA&&2018 Jun 2.34329516&390&2018 Jun 2.50095916&220&2018 Jun 2.50095916&220\\[-5pt]
2019 MO&&2019 Jun 22.32928515&576&2019 Jun 22.42462415&480&2019 Jun 22.42462415&480\\[-5pt]
2022 EB5&&2022 Mar 11.80847610&115&2022 Mar 11.88684010&12&2022 Mar 11.88684010&12\\[-5pt]
2022 WJ1&&2022 Nov 19.20348201&128&2022 Nov 19.32745602&27&2022 Nov 19.32745602&27\\[-5pt]
2023 CX1&&2023 Feb 12.84591707&233&2023 Feb 13.11952707&11&2023 Feb 13.11952707&11\\
\hline
\multicolumn2l{Fly-bys}\\
\hline
2008 EK68&0.0000000&2008 Mar 5.44060&1622&2008 Mar 5.48066&1650&2008 Mar 5.44060&1622\\[-5pt]
2011 AE3&0.0000000&2011 Jan 4.23829&936&2011 Jan 4.28106&946&2011 Jan 4.23829&936\\[-5pt]
2016 QY84&0.0000000&2016 Aug 29.28386&2206&2016 Aug 29.40878&2241&2016 Aug 29.28386&2206\\[-5pt]
2011 CH22&0.0000000&2011 Feb 7.52627&1737&2011 Feb 7.54497&1743&2011 Feb 7.52627&1737\\[-5pt]
2017 UL52&0.0000000&2017 Oct 21.30085&1627&2017 Oct 21.47499&1659&2017 Oct 21.30085&1627\\[-5pt]
2010 XC&0.0000000&2010 Dec 1.30659&2731&2010 Dec 1.36495&2710&2010 Dec 1.36495&2710\\[-5pt]
2012 HA34&0.0000000&2012 Apr 30.37929&2402&2012 Apr 30.98138&1875&2012 Apr 30.98138&1875\\[-5pt]
2009 JE1&0.0000000&2009 May 4.36574&4118&2009 May 4.43411&4079&2009 May 4.43411&4079\\[-5pt]
2008 JD33&0.0000000&2008 May 15.28689&5061&2008 May 15.39872&5088&2008 May 15.28689&5061\\[-5pt]
2010 SV15&0.0000000&2010 Sep 30.27850&8429&2010 Sep 30.37866&8492&2010 Sep 30.27850&8429\\[-5pt]
2012 TC4&0.0000001&2012 Oct 4.46766123&4420&2017 Dec 14.30066710&30353&2017 Oct 12.16121419&65\\[-5pt]
2017 VL2&0.0000008&2017 Nov 10.47521&787&2017 Nov 26.34580&12206&2017 Nov 10.47521&787\\[-5pt]
2021 VD8&0.0000010&2021 Nov 10.27772302&719&2021 Nov 12.55660003&3022&2021 Nov 10.27772302&719\\[-5pt]
2022 GX2&0.0000018&2022 Apr 2.31649411&2350&2022 Apr 3.17817111&3062&2022 Apr 2.31649411&2350\\[-5pt]
2021 XK1&0.0000034&2021 Dec 2.34278105&4959&2022 Jan 9.35733004&36101&2021 Dec 2.34278105&4959\\[-5pt]
2021 AY5&0.0000100&2021 Jan 12.50387510&9245&2021 Feb 18.28549&23557&2021 Jan 12.50387510&9245\\[-5pt]
2020 AP1&0.0000100&2020 Jan 4.37133&879&2020 Jan 5.44163&1359&2020 Jan 4.37133&879\\[-5pt]
2014 WE6&0.0000100&2014 Nov 17.19435&1358&2014 Nov 21.46668602&3001&2014 Nov 17.19435&1358\\[-5pt]
2007 XO&0.0000100&2007 Dec 4.27109&7549&2007 Dec 4.45258&7242&2007 Dec 4.45258&7242\\[-5pt]
2002 EM7&0.0000100&2002 Mar 12.30675&3805&2002 Apr 6.26806&27112&2002 Mar 12.30675&3805\\[-5pt]
2004 FU162&0.0000200&2004 Mar 31.27744&357&2004 Mar 31.30799&329&2004 Mar 31.30799&329\\[-5pt]
2022 BN2&0.0000200&2022 Jan 27.38223110&1271&2022 Jan 28.42156911&508&2022 Jan 28.42156911&508\\[-5pt]
2016 AH164&0.0000200&2016 Jan 13.31227&696&2016 Jan 18.28648808&3592&2016 Jan 13.31227&696\\[-5pt]
2014 HB177&0.0000200&2014 Apr 29.36076&3782&2014 May 5.03469&850&2014 May 5.03469&850\\[-5pt]
2020 SP6&0.0000200&2020 Sep 28.20162122&2168&2020 Nov 18.27896803&43572&2020 Sep 28.20162122&2168\\[-5pt]
2015 KE&0.0000300&2015 May 18.20995&4004&2016 Aug 25.24826&6646&2015 May 18.20995&4004\\[-5pt]
2018 WG2&0.0000300&2018 Nov 29.23915&904&2018 Nov 30.73289522&203&2018 Nov 30.73289522&203\\[-5pt]
2021 XA6&0.0000300&2021 Dec 12.25166822&537&2022 Jan 2.87397&10521&2021 Dec 12.25166822&537\\[-5pt]
2018 VP1&0.0000400&2018 Nov 3.27249502&451&2018 Nov 16.24026503&11303&2018 Nov 3.27249502&451\\[-5pt]
2020 HC11&0.0000400&2020 Apr 17.36457314&62789&2020 Sep 17.59246403&72547&2020 May 29.17815&29435\\[-5pt]
2018 XQ2&0.0000400&2018 Dec 10.24230&3675&2019 Jan 7.06764101&9841&2018 Dec 10.24230&3675\\[-5pt]
2020 VT4&0.0000500&2020 Nov 14.32062802&428&2020 Nov 19.26649&3814&2020 Nov 14.32062802&428\\[-5pt]
2018 DN4&0.0000500&2018 Feb 26.24992&2524&2018 Feb 26.39447&2748&2018 Feb 26.24992&2524\\[-5pt]
2017 UR2&0.0000500&2017 Oct 19.39518&1620&2017 Oct 21.26632&3343&2017 Oct 19.39518&1620\\[-5pt]
2017 TU1&0.0000500&2017 Oct 1.41139702&11127&2017 Oct 14.10575&2074&2017 Oct 14.10313&2074\\[-5pt]
2011 GP28&0.0000500&2011 Apr 4.36451&3089&2011 Apr 5.33846&1864&2011 Apr 5.33846&1864\\[-5pt]
2020 QG&0.0000600&2020 Aug 16.43273200&198&2020 Aug 18.00587000&1317&2020 Aug 16.43273200&198\\[-5pt]
2021 UA1&0.0000600&2021 Oct 25.25544801&147&2021 Oct 26.11606101&1110&2021 Oct 25.25544801&147\\[-5pt]
2014 LY21&0.0000600&2014 Jun 2.41612&1345&2014 Jun 2.46118&1300&2014 Jun 2.46118&1300\\[-5pt]
2017 LD&0.0000600&2017 May 16.29848813&8555&2017 Jun 30.35601401&9991&2017 Jun 5.40331&1151\\[-5pt]
2010 VP139&0.0000600&2010 Nov 14.27911&1688&2010 Nov 14.32499&1723&2010 Nov 14.27911&1688\\[-5pt]
2009 VZ39&0.0000600&2009 Nov 10.27204&1674&2009 Nov 10.31546&1646&2009 Nov 10.31546&1646\\[-5pt]
2020 GB1&0.0000600&2020 Apr 2.53492414&3205&2020 Apr 6.96972807&413&2020 Apr 6.96972807&413\\[-5pt]
2020 RD4&0.0000700&2020 Sep 12.49828100&2017&2020 Sep 14.24158823&538&2020 Sep 14.24158823&538\\[-5pt]
2022 KZ&0.0000700&2022 May 21.45054&4504&2022 May 25.45400520&978&2022 May 25.45400520&978\\[-5pt]
2022 EQ&0.0000700&2022 Mar 2.32776010&852&2022 Mar 2.98145208&250&2022 Mar 2.98145208&250\\[-5pt]
2023 BU&0.0000700&2023 Jan 21.34299109&1376&2023 Jan 31.91281115&1197&2023 Jan 26.96052107&31\\
\enddata
\tablecomments{The seven telescopically detected Earth-impactors and 47 Earth fly-bys with minimal geocentric distances less than 0.00008 au as provided by NEO Earth Closes Approaches Web site\footnote{\url{https://cneos.jpl.nasa.gov/ca/}} with observation times provide by the astroquery.mpc\footnote{\url{https://astroquery.readthedocs.io/en/latest/mpc/mpc.html\#}} Python package.  The epoch and geocentric distance in kilometres for the initial observation, the last observation, and the closest observation are arrived at by interpolating ephemerides for the provided objects acquired from the JPL Horizons Telnet ephemeris service\footnote{telnet horizons.jpl.nasa.gov 6775}.}  
\end{deluxetable*}

Though the USG object was telescopically detected when unusually close and small, the event does refuse superlative labelling, though barely.  At the time of impact, the object was only the fifth Earth-impacting object to be detected in space.  Subsequent to 2020 September 18, two additional impacting objects, 2022 EB5\footnote{\url{http://www.cbat.eps.harvard.edu/iau/cbet/005100/CBET005108.txt}} and 2022 WJ1\footnote{\url{https://www.jpl.nasa.gov/news/nasa-program-predicted-impact-of-small-asteroid-over-ontario-canada}} were observed in space prior to impact on 2022 March 11 and 2022 November 19 repectively.  At the time of the USG event, the ATLAS observation was the closest observation of a natural preatmospheric object.  
This proximity analysis was performed using the Python astroquery.mpc\footnote{\url{https://astroquery.readthedocs.io/en/latest/mpc/mpc.html\#}} package to download all observation times of the six Earth-impacting object observations as well as the 47 near-Earth fly-bys whose minimal geocentric distance was less than that of the USG object (see Table \ref{tab:nearObs}).  Then using the JPL Horizons Telnet ephemeris system\footnote{telnet horizons.jpl.nasa.gov 6775} we acquired the nominal geocentric distance for each object observation.  Two objects  having a telescopic observation distance of less than the USG object's 11920 km geocentric distance are 2022 EB5 at a calculated 11854 km and 2023 CX1 at 11125 km.   Prior to 2020 September 18, the closest object observation was that of 2018 LA at approximately 22000 km.  A superlative that does hold for the ATLAS image of the USG event is that it is the closest {\it initial} observation of a preatmospheric object, far exceeding 2022 EB5 at 115,000~km and 2021 UA1 at 147000~km.

By no means is the USG object the smallest object observed in space.  The MPCORB\footnote{\url{https://www.minorplanetcenter.net/iau/MPCORB/MPCORB.DAT}},
NEODyS\footnote{\url{https://newton.spacedys.com/~neodys2/neodys.cat}}$^{,}$\footnote{\url{https://newton.spacedys.com/~neodys2/neodys.ctc}}
and JPL-SSD Small Body\footnote{\url{https://ssd.jpl.nasa.gov/dat/ELEMENTS.UNNUM}} asteroid and NEA databases all list 2008 TS26 at absolute magnitude 33.1 - 33.2, and 2021 BO at 32.9. They both exceed the estimated H=32.5 of the USG event, even without regard to the disproportionately high $H$ due to an exceedingly low albedo.

From the point of view of the FROSTI project and the hope to identify preatmospheric images of objects having produced fireballs, the more than 4$\degree$ discrepancy in object location on the image plane is a matter of concern.  We were indeed fortunate that the predicted location of the object and the actual observation of the object were both on the same image, aided by the wide 5.5$\degree$ field of ATLAS. Our underestimate of the real uncertainties of USG events could result in images with fireballs going unflagged by FROSTI, particularly for sky survey images with small fields.  In like scenarios to this event, using standard deviations equal to the USG most significant reported digit appears to underestimate uncertainties.  However, a wholesale increase of uncertainty cloud size for all USG events within FROSTI will also generate many false-positive results.  Granted that the 2020 September 18 detection was somewhat unique given the proximity of the object, thus minimizing the extent of in-sky positional uncertainty.  However,  experience with visualizing uncertainty cloud behaviour as one moves backward in time prior to an impact shows that the decreasing angular size of the cloud due to increased distance and the divergence due to increasing clone spread tend to cancel each other out.  Although CNEOS provides data on bolides that would go otherwise optically unobserved, this case study further supports the general result found by \citet{Devillepoix2019}, namely that the USG derived trajectories are much less precise than those derived from ground-based fireball observations.

\section{Summary} \label{sec:summary}

The USG 20200918 fireball was the first successful identification from archival imagery of a serendipitous preatmospheric image of an observed fireball. We have shown that the ATLAS detection of a streak just ten minutes prior to impact is broadly consistent with the CNEOS listed USG state vector within its large expected uncertainties \citep{Devillepoix2019}, while ruling out confusion from known satellites or an unassociated natural object. Given the large USG uncertainties, the detection via the FROSTI survey was very fortuituous: imaging so near to impact the large uncertainty was still smaller than the wide ATLAS field of view.

Our analysis of the fireball produces a preferred estimate of 0.4 kt TNT total energy. This is based on a synthesis of the CNEOS and GLM lightcurve and infrasound source energies. This energy, together with a pre-atmosphere speed of 12.7 km s$^{-1}$ measured by combining the USGmstate vector suggest a mass of order 23~t. Through comparison with 2008~TC3, which entered at a similar speed and behaved similarly in atmosphere based on the recorded fireball lightcurve, we suggest the most likely bulk density is also around 1600 kg m$^{-3}$ and associated diameter of 3 m. The telescopic lightcurve amplitude suggests an axis ratio of 1.6 while lightcurve periodicity is consistent with a rotation period of about 16 seconds. 

The resulting best-fit orbit results in a high probability of escape from the $\nu_6$ secular resonance. The apparent magnitude and size are most consistent with a very low albedo C-complex object. Whether the apparent structural weakness is due to a rubble-pile structure or microporosity is unclear.

The event has demonstrated that the the FROSTI project can add to the collection of in-space imaged Earth impactors, collection of which was previously dependent on pre-contact object detection.  The ATLAS image is the closest-ever initial observation of a preatmospheric object, and rivals the observations of 2022 EB5 and 2023 CX1 as one of the closest observations of any such object.

Taken on its own, the absence of reported uncertainties for the USG calculated object trajectory makes its interpretation more difficult, but was sufficiently accurate in this case to allow us to identify a historical ATLAS image of the object.

\section{Acknowledgements} \label{sec:acknowledgments}

This work was supported in part by the NASA Meteoroid Environment Office under cooperative agreement 80NSSC21M0073 and by the Natural Sciences and Engineering Research Council of Canada (Grants no. RGPIN-2018-05659), and by the Canada Research Chairs Program.

This work has made use of data from the Asteroid Terrestrial-impact
  Last Alert System (ATLAS) project. ATLAS is primarily funded to search
  for near earth asteroids through NASA grants NN12AR55G, 80NSSC18K0284,
  and 80NSSC18K1575; byproducts of the NEO search include images and
  catalogs from the survey area.  The ATLAS science products have been
  made possible through the contributions of the University of Hawaii
  Institute for Astronomy, the Queen's University Belfast, the Space
  Telescope Science Institute, the South African Astronomical Observatory (SAAO),
  and the Millennium Institute of Astrophysics (MAS), Chile.

 We thank the many contributors to the FROSTI image database used for image searching, including \textbf{ATLAS}, Catalina, CNEOST, DeCAM, LONEOS, Pan-STARRS, WISE and NEOWISE, ZTF, the Minor Planet Center and its contributing observatories, and the Canadian Astronomy Data Centre for CFHT and HST data
  
  We thank J. Borovi\v{c}ka for helpful discussions. 

  We also thank the reviewers for their suggestions leading to several notable improvements to the article.


\pagebreak
\bibliography{CNEOS2020-09-18}{}
\bibliographystyle{aasjournal}



\end{document}